\documentclass[useAMS,usenatbib,usegraphicx]{mn2e}
\usepackage{color}
\usepackage{psfig}
\usepackage{amssymb}
\usepackage{savesym}
\usepackage{multicol}
\usepackage{multirow}
\usepackage{supertabular}

\newcommand{\aj}{AJ}
\newcommand{\apj}{ApJ}
\newcommand{\mnras}{MNRAS}

\newcommand{\apjl}{ApJLetters}
\newcommand{\aap}{A\&A}

\newcommand{\araa}{ARA\&A}
\newcommand{\nat}{Nature}
\newcommand{\pasa}{PASA} 
\newcommand{\NS}{{\sc NewSink}}
\newcommand{\NSs}{{\sc NewSinks}}
\newcommand{\NSg}{{\sc NewSink} }
\newcommand{\NSsg}{{\sc NewSinks} }
\newcommand{\OS}{{\sc OldSink}}
\newcommand{\OSs}{{\sc OldSinks}}
\newcommand{\OSg}{{\sc OldSink} }
\newcommand{\OSsg}{{\sc OldSinks} }
\newcommand{\US}{{\sc UrSink}}
\newcommand{\USs}{{\sc UrSinks}}
\newcommand{\USg}{{\sc UrSink} }
\newcommand{\USsg}{{\sc UrSinks} }
\newcommand{\dsh}{\mbox{---}}
\newcommand{\refrpt}{}
\title[An improved sink particle algorithm for SPH simulations]{An improved sink particle algorithm for SPH simulations}
\author[Hubber, Walch and Whitworth]{D. A. Hubber$^{1,2}$, S. Walch$^{3,4}$, A. P. Whitworth$^3$ \\
$^{1}$Department of Physics and Astronomy, University of Sheffield, Hicks Building, Hounsfield Road, Sheffield, S3 7RH, UK \\
$^{2}$School of Physics and Astronomy, University of Leeds, Leeds, LS2 9JT, UK \\
$^{3}$School of Physics and Astronomy, Cardiff University, Queens Buildings, The Parade, CF24 3AA, UK \\
$^{4}$Max-Planck-Institut f{\" u}r Astrophysik, Karl-Schwarzschild-Str. 1, Garching, D84758, Germany}

\begin{document}
\date{December 4th, 2012}
\pagerange{\pageref{firstpage}--\pageref{lastpage}} \pubyear{2012}
\label{firstpage}
\maketitle

\begin{abstract}
Numerical simulations of star formation frequently rely on the implementation of sink particles, (a) to avoid expending computational resource on the detailed internal physics of individual collapsing protostars, (b) to derive mass functions, binary statistics and clustering kinematics (and hence to make comparisons with observation), and (c) to model radiative and mechanical feedback; sink particles are also used in other contexts, for example to represent accreting black holes in galactic nuclei. We present a new algorithm for creating and evolving sink particles in SPH simulations, which appears to represent a significant improvement over existing algorithms {\refrpt -- particularly in situations where sinks are introduced after the gas has become optically thick to its own cooling radiation and started to heat up by adiabatic compression}. (i) It avoids spurious creation of sinks. (ii) It regulates the accretion of matter onto a sink so as to mitigate non-physical perturbations in the vicinity of the sink. (iii) Sinks accrete matter, but the associated angular momentum is transferred back to the surrounding medium. With the new algorithm -- and modulo the need to invoke sufficient resolution to capture the physics preceding sink formation -- the properties of sinks formed in simulations are essentially independent of the user-defined parameters of sink creation, or the number of SPH particles used.
\end{abstract}

\section{Introduction} \label{SEC:INTRO}%

Star formation is a collective process: many stars are born in multiple systems \citep[e.g.][]{DM1991}, and most are born in clusters \citep[e.g.][]{LL2003}. Consequently, a realistic simulation that captures the important interactions between neighbouring protostars, as they form, is only feasible if some of the detailed internal physics of individual protostars is sacrificed. Otherwise all the available computational resource is commandeered by the first protostar to form, and events in the rest of the computational domain grind to a halt. It is for this reason that Lagrangian sink particles were first developed by \citet{Bate1995}. Subsequently they have also been used to model single accreting objects, for example black holes in galactic nuclei, where the focus is not on the detailed hydrodynamics in the immediate vicinity of the black hole, but on feedback from the black hole on much larger scales \citep[e.g.][]{Springel2005}.

In the most basic implementation, a sink is created when and where some lump in the computational domain looks like it will inevitably condense out gravitationally (into a single star, or a tight multiple system, or a black hole). Sinks can also be introduced into the initial conditions for a simulation, to represent pre-existing condensations (stars or black holes). A basic sink is a point mass, and therefore it interacts with the surrounding matter gravitationally. Once created, a sink can grow by accreting nearby matter that looks like it would inevitably become part of the same condensation. Even at this most basic level, there are issues with how to identify lumps that should be converted into sinks, and how to determine which matter they should subsequently accrete.

At a higher level of sophistication, sink properties (masses, positions, velocities, accretion rates) are used to infer stellar mass functions, binary statistics, cluster kinematics and {\refrpt the star formation rate in turbulent molecular clouds} \citep[e.g.][]{Bonnell2001, Kits+Whit2002, Bateetal2003, Goodwin2004, Goodwin2004b, Bonnell2004, Goodwin2006, Bonnell2006, Bate2009, Attwood2009, Stametal2011, Walch2012, FK2012}. They can also be used to model radiative and mechanical feedback from protostars, using simple models or phenomenological prescriptions for the physics inside the sink \citep[e.g.][]{Stametal2005, Dale2005,  Krumholz2007, Stametal2009b, Offner2009, Bisbas2009, Bate2010, Peters2010, Wang2010, Dale2011, Peters2011, Bate2012, Cunningham2011, Dale2012}.

In cosmological simulations, sink particles are used to model the formation and growth of black holes, and hence AGN feedback \citep{Springel2005, DiMatteo2005, DiMatteo2008, Johansson2009, Debuhr2010, Debuhr2011, Choi2012}. For example, in \citet{Springel2005} the efficiency of AGN feedback is estimated using the Bondi-Hoyle-Littleton theory \citep{Hoyle1939, Bondi1944, Bondi1952} acting on scales of a few tens of parsecs, whereas the actual Schwarzschild radius of a typical supermassive black hole is of order a few solar radii.

Two main algorithms are currently used to model star formation, Smoothed Particle Hydrodynamics \citep[SPH;][]{Lucy1977, Gingold1977} and Adaptive Mesh Refinement \citep[AMR;][]{Berger1989, Berger1984, Dezeeuw1993, MacNeice2000}. This paper is concerned with sinks in SPH.

In SPH simulations of star formation, two requirements are critical. First, the mass of a single SPH particle, $m_{_{\rm SPH}}$, should be chosen \citep[e.g.][]{BateBurkert1997, Whit1998, Attwood2009, Bate2009, Stamatellos2009a}, or adjusted \citep[e.g.][]{Kits+Whit2002}, so that the Jeans mass is always resolved. Second, sink creation should be considered only if the density exceeds a user-defined threshold, $\rho_{_{\rm SINK}}$, and $\rho_{_{\rm SINK}}$ should in turn exceed the density at which the minimum Jeans mass obtains, so that the code can capture properly the full range of fragmentation, and the formation of first cores. {\refrpt Thus our main concern here is with sinks created from gas that is already being heated by adiabatic compression; for a discussion of how to treat sinks created at lower densities from approximately isothermal gas the reader is refered to \citet{Federrath2010}.} For contemporary local star formation, involving molecular gas at $T\!\sim\! 10\,{\rm K}$, these requirements reduce to $m_{_{\rm SPH}}\la 10^{-5}\,{\rm M}_{_\odot}$ and $\rho_{_{\rm SINK}}\gg 10^{-13}\,{\rm g}\,{\rm cm}^{-3}$. Once the density exceeds  $\rho_{_{\rm SINK}}$, the criteria that are applied to determine (i) whether a sink actually is created, (ii) what matter it immediately assimilates, and (iii) what matter it subsequently accretes, must be formulated so as to obtain close correspondence between the properties of the sinks that do form, and the stars that should have formed, under the given physical circumstances.

In AMR simulations of star formation, the computational domain is divided adaptively into finer meshes -- and as appropriate these meshes are subsequently de-constructed -- so as to deliver high resolution only when and where it is needed. There is a minimum mesh spacing, $\ell_{_{\rm MIN}}$, below which refinement is prohibited. In the original implementation of sink particles in AMR \citep{Krumholz2004}, a sink is introduced wherever the local Jeans length falls below $4\ell_{_{\rm MIN}}$ and is therefore no longer adequately resolved. There are then rules to determine the rate at which mass is accreted onto the sink from the surrounding cells, and to identify the circumstances under which neighbouring sinks are merged. The angular momentum of the accreted mass is left behind in the cell from which it is accreted. \citet{Offner2008} have shown that this implementation is quite robust, in the sense that there is good correspondence between the properties of sinks formed in simulations having the same initial conditions but different $\ell_{_{\rm MIN}}$. To capture the full range of fragmentation, and the formation of first cores, requires $\ell_{_{\rm MIN}}\ll3\,{\rm AU}$, for contemporary local star formation. {\refrpt We note that in some AMR sink implementations \citep[e.g.][]{Federrath2010} the angular momentum of matter accreted by the sink is not left behind, but is assimilated by the sink.}

Recently, \citet{Federrath2010} have simulated the formation of a star cluster with SPH and AMR, using comparable resolution, and have developed a set of sink-creation criteria which ensure that the two methods give very similar results, in terms of the masses, locations and accretion histories of the sinks formed. However, {\refrpt although these simulations investigate the consequences of different sink  creation criteria, they do not explore in detail} the dependence of the sink properties on the resolution, or on the user-defined parameters of sink creation. The present paper addresses these issues.

In Section \ref{SEC:NEWSINKS} we describe in detail a new algorithm for the creation and evolution of sinks, explaining the motivation for each of its features; we call these \NSs. In Sections \ref{SEC:OLDSINKS} and \ref{SEC:URSINKS}, we describe the sink algorithms that have been used previously in SPH simulations of star formation -- hereafter {\it standard sinks} -- focussing on the particular implementations that we use for comparison tests. Since all but one of the new features in \NSsg relate to their evolution, rather than their creation, Section \ref{SEC:OLDSINKS} defines \OSs, which are created in the same way as \NSs, but evolved in the manner of standard sinks, thereby allowing us to isolate those aspects of \NSsg that have to do purely with sink evolution. On the other hand, since most extant SPH simulations of star formation that invoke sinks actually use a less robust algorithm to create sinks than that used by \NSsg and \OSs, Section \ref{SEC:URSINKS} defines \USs, which are created {\it and} evolved in the manner of standard sinks. The basic properties of the different types of sink are summarised in Table \ref{TAB:SINKTYPES}. Section \ref{SEC:TESTS} presents the results of tests performed with the different types of sink to demonstrate that when \NSsg are used the results obtained are only weakly dependent on the resolution and the user-defined parameters of sink-creation, whereas when \OSsg are used the results are strongly dependent and not converged, and when \USsg are used the results are divergent. General aspects of the different types of sink are discussed in Section \ref{SEC:DISC}. Our main conclusions are summarised in Section \ref{SEC:CONC}.

\begin{table*}
\begin{center}
\begin{tabular}{llr|c|ccc}\hline
\multicolumn{3}{r}{\sc Type of Sink}$\;{\bf \longrightarrow}\;\,$                                                                     &
  $\hspace{1cm}$ & \NS              & \OS              & \US              \\\hline
{\sc Creation criteria:}            & {\sc Equation:}                                              &                        &
                 &                  &                  &                  \\
$\;\;\;\;\;$density                 & $\;\;\;\;\;\rho_i\!>\!\rho_{_{\rm SINK}}$                    & (\ref{EQN:RHOCRIT})    &
                 & $\bigstar$       & $\bigstar$       & $\bigstar$       \\
$\;\;\;\;\;$overlap                 & $\;\;\;\;\;|{\bf r}_i-{\bf r}_s|\!>\!X_{_{\rm SINK}}h_i+R_s$ & (\ref{EQN:VRLPCRIT})   &
                 & $\bigstar$       & $\bigstar$       & $\bigstar$       \\
$\;\;\;\;\;$potential minimum       & $\;\;\;\;\;\phi_i\!<\!\mbox{\sc min}\left\{\phi_j\right\}$   & (\ref{EQN:POTMINCRIT}) &
                 & $\bigstar$       & $\bigstar$       & \dsh             \\
$\;\;\;\;\;$Hill Sphere             & $\;\;\;\;\;\rho_i\!>\!\rho_{_{\rm HILL}}\!=\,...$            & (\ref{EQN:HILLCRIT})   &
                 & $\bigstar$       & $\bigstar$       & \dsh             \\
$\;\;\;\;\;$acceleration divergence & $\;\;\;\;\;(\nabla\!\cdot\!{\bf a})_i\!<\!0$                 & (\ref{EQN:DIVACCCRIT}) &
                 & \dsh             & \dsh             & $\bigstar$       \\
$\;\;\;\;\;$velocity divergence     & $\;\;\;\;\;(\nabla\!\cdot\!{\bf v})_i\!<\!0$                 & (\ref{EQN:DIVVELCRIT}) &
                 & \dsh             & \dsh             & $\bigstar$       \\
$\;\;\;\;\;$non-thermal energy      & $\;\;\;\;\;E_{_{\rm NT}}\!<\!0$                              & (\ref{EQN:ENERGYCRIT}) &
                 & \dsh             & \dsh             & $\bigstar$       \\\hline
{\sc Structure:}                    & {\sc Evolution:}                                             &                        &
                 &                  &                  &                  \\
$\;\;\;\;\;$interaction-zone        & $\;\;\;\;\;$regulated accretion                              &                        &
                 & $\blacksquare$   & \dsh             & \dsh             \\
$\;\;\;\;\;$exclusion-zone          & $\;\;\;\;\;$instantaneous accretion                          &                        &
                 & \dsh             & $\blacksquare$   & $\blacksquare$   \\\hline
\end{tabular}
\caption{The creation criteria, intrinsic structures, and evolution procedures for the different types of sink. Column 1 gives the various creation criteria and the two different intrinsic structures. Column 2 gives equations for the various creation criteria, and the two different modes of evolution. Column 3 gives equation numbers. In columns 4 (\NS), 5 (\OS) and 6 (\US), stars indicate which creation criteria are invoked, and filled squares indicate which structure/evolution combination is used. Relative to \NSs, \OSsg have the same creation criteria, but different evolution procedures; \USsg have different creation criteria {\it and} different evolution procedures. Most extant SPH simulations of star formation have been performed with standard sinks like \USs.}
\end{center}
\label{TAB:SINKTYPES}
\end{table*}

\section{The \NSg  algorithm} \label{SEC:NEWSINKS}%

Here we describe and justify the procedures used to define the internal structure and translation of a \NSg (Section \ref{SEC:NEWSTRUCT}), to trigger the creation of a \NSg (Section \ref{SEC:NEWCREATE}), to accrete matter onto an existing \NSg (Section \ref{SEC:NEWACCRETE}), to update the properties of a \NSg (Section \ref{SEC:NEWUPDATE}), and to pass the angular momentum acquired by a \NSg back to the surrounding matter (Section \ref{SEC:NEWANGMOM}).

\subsection{Internal structure and advection of a \NS}\label{SEC:NEWSTRUCT}%

A \NSg comprises a central point mass (hereafter, the {\it point-mass}) and a concentric spherical volume of radius $R_s$ (hereafter, the {\it interaction-zone}) that moves with, and remains centred on, the point-mass. When we refer to the mass, $M_s$, position, ${\bf r}_s$, velocity, ${\bf v}_s$, or angular momentum, ${\bf L}_s$, of a \NS, we mean the mass, position, velocity and spin angular momentum of its point-mass. Once created, a \NSg is tracked with the same integration routine as an SPH particle, modulo that it only experiences gravitational acceleration. The interaction-zone contains live SPH particles that have not yet been assimilated by the point-mass, and which are intended to represent the continuation of any accretion disc, inflow stream, or other coherent structure immediately outside the interaction-zone. The interaction-zone is one of the key features that distinguishes a \NSg from a standard sink.

\subsection{Criteria for creation of a \NS} \label{SEC:NEWCREATE}%

The creation of a \NSg is triggered by an SPH particle. Specifically, if SPH particle $i$, having smoothing length $h_i$, satisfies the four criteria below, it is replaced with a \NS, $s$. At the moment of creation, the point-mass of $s$ has the same mass, position and velocity as $i$ did, and the radius of its interaction-zone is
\begin{eqnarray}\label{EQN:SINKRAD}
R_s&=&X_{_{\rm SINK}}h_i\,. 
\end{eqnarray}
$X_{_{\rm SINK}}$ is a user defined parameter, chosen so that the neighbours of $i$ all fall inside the \NSg at the moment of its creation. Thus, for the M4 kernel, we advocate $X_{_{\rm SINK}}=2$. By adopting a larger $X_{_{\rm SINK}}$ one obtains smoother accretion onto a \NS, but at the expense of coarser resolution.

\subsubsection{Density criterion} \label{SEC:RHOCRIT}%

The first criterion for SPH particle $i$ to trigger the creation of a \NSg is that the density at its location, $\rho_i\equiv\rho_{_{\rm SPH}}({\bf r}_i)$, should exceed a user-defined threshold, $\rho_{_{\rm SINK}}$, i.e.
\begin{equation}\label{EQN:RHOCRIT}
\rho_i > \rho_{_{\rm SINK}}\,.
\end{equation}
We advocate $\rho_{_{\rm SINK}}\geq 10^{-11}\,{\rm g}\,{\rm cm}^{-3}$, so that \NSsg normally form in condensations that are well into their Kelvin-Helmholtz contraction phase.

\subsubsection{Sink-overlap criterion} \label{SEC:VRLPCRIT}%

The second criterion for SPH particle $i$ to trigger the formation of a \NSg is that it should not, at the moment of creation, overlap another \NS, i.e.
\begin{eqnarray}\label{EQN:VRLPCRIT}
|{\bf r}_i-{\bf r}_{s'}|&>&X_{_{\rm SINK}}h_i\,+\,R_{s'}\,,
\end{eqnarray}
for all pre-existing \NSs, $s'$.

\subsubsection{Gravitational potential minimum criterion} \label{SEC:POTMINCRIT}%

The third criterion for SPH particle $i$ to trigger the formation of a \NSg is that its gravitational potential, $\phi_i$, should be lower than that of all it neighbours, $j$, i.e.
\begin{eqnarray}\label{EQN:POTMINCRIT}
\phi_i&<&\mbox{\sc min}\left\{\phi_j\right\}\,.
\end{eqnarray}
This criterion, introduced by \citet{Federrath2010}, ensures that \NSsg are only created near resolved potential minima (and hence, presumably, resolved density peaks).

\subsubsection{Hill Sphere criterion} \label{SEC:HILLCRIT}%

The fourth criterion for SPH particle $i$ to trigger the formation of a \NSg is that its density should satisfy
\begin{eqnarray}\label{EQN:HILLCRIT}
\rho_i&>&\rho_{_{\rm HILL}}\;\,\equiv\;\,\frac{3\,X_{_{\rm HILL}}\,\left(-\Delta {\bf r}_{is'}\!\cdot\!\Delta {\bf a}_{is'}\right)}{4\,\pi\,G\,|\Delta {\bf r}_{is'}|^2}\,,
\end{eqnarray}
for all pre-existing \NSs, $s'$. Here $X_{_{\rm HILL}}$ is a user-defined parameter with default value $X_{_{\rm HILL}}=4$, $\Delta {\bf r}_{is'}\equiv{\bf r}_i-{\bf r}_{s'}$ and $\Delta {\bf a}_{is'}\equiv{\bf a}_i-{\bf a}_{s'}$ are the position and acceleration of particle $i$ relative to $s'$.\footnote{Here, and in the sequel, we adopt the convention that $\Delta{\bf q}_{ab}\equiv{\bf q}_a-{\bf q}_b$ is the difference in some quantity ${\bf q}$ between two particles $a$ and $b$, evaluated at the same time; thus, for example, $\Delta{\bf r}_{ij}\equiv{\bf r}_i-{\bf r}_j$ is the instantaneous position of SPH particle $i$ relative to SPH particle $j$. In contrast, we adopt the convention that $\delta{\bf q}_c$ is the change in some quantity ${\bf q}_c$ associated with particle $c$, due to the evolution of the system; thus, for example, $\delta{\bf L}_s$ is the increment to the angular momentum of sink $s$, and $\delta t_s$ is the timestep of $s$.} This criterion deals with the situation where a condensation undergoing Kelvin-Helmholtz contraction has a \NS $\;s'$ in its interior, but is much more extended than that \NS. Eqn. (\ref{EQN:HILLCRIT}) ensures that a second \NS, seeded by SPH particle $i$, can form in the outskirts of the condensation only if there is a density peak at ${\bf r}_i$ that dominates the local gravitational field.

\subsection{Criteria for accretion by a \NS} \label{SEC:NEWACCRETE}%

\subsubsection{The SPH particles inside a \NS}\label{SEC:THELIST}%

Under normal circumstances, SPH particles that enter the interaction-zone of a \NSg are not accreted immediately by its point-mass. In the first instance, they are simply added to a list of the SPH particles inside the interaction-zone, its interaction-list. These SPH particles may subsequently be accreted by the point-mass, possibly over several timesteps, but they may leave the interaction-zone before this happens. If an SPH particle finds itself inside the interaction-zones of more than one existing \NS, it is added to the interaction-list of the \NSg whose point-mass is closest. In this context, it is appropriate to note that, although a \NSg is never created overlapping another \NS, its motion may thereafter lead to an overlap. We do not allow sinks to merge \citep[cf.][]{Krumholz2004}.

To determine whether, and how quickly, the SPH particles in the interaction-zone of an existing \NSg are assimilated by its point-mass, we consider the limiting cases of spherically-symmetric radial accretion, and disc accretion.

\subsubsection{Timescale for spherically symmetric radial accretion} \label{SEC:tRAD}%

In spherical symmetry, the rate at which mass flows inwards across a spherical surface at radius $r$ is
\begin{eqnarray}
\dot{M}(r)&=&\,-\;4\pi r^2\rho(r)v_{_{\rm RAD}}(r)\,.
\end{eqnarray}
At the position of an SPH particle $j$, in the interaction-zone of \NSg $\,s$, this inflow rate becomes
\begin{eqnarray}
\dot{M}_j&=&\,-\;4\pi\,|\Delta{\bf r}_{js}|\,\Delta{\bf r}_{js}\!\cdot\!\Delta{\bf v}_{js}\,\rho_j\,.
\end{eqnarray}
The timescale for radial accretion is obtained by dividing the net mass of the SPH particles in the interaction-zone by a weighted sum of the inflow rates:
\begin{eqnarray}\label{EQN:tRAD}
\left<t_{_{\rm RAD}}\right>_s\!\!&\!\!=\!\!&\!\!\frac{\sum_j\!\left\{m_j\right\}\,{\cal W}}{4\pi\sum\limits_{j}\!\left\{|\Delta{\bf r}_{js}|\Delta{\bf r}_{js}\!\cdot\!\Delta{\bf v}_{js}m_jW(|\Delta{\bf r}_{js}|,H_s)\right\}},\\
{\cal W}&=&\sum_j\!\left\{m_jW(|\Delta{\bf r}_{js}|,H_s)/\rho_j\right\}\,.
\end{eqnarray}
The weighting here uses the kernel-function, $W$. The smoothing length of the sink, $H_s$, is adjusted so that the extent of the kernel function equals $R_s$; for example, with the M4 kernel, $H_s\!=\!R_s/2$. Consequently SPH particles in the outer regions of the interaction-zone make a smaller contribution than those closer to the point-mass. {\refrpt ${\cal W}$ ensures that the sum is accurately normalised.}

\subsubsection{Timescale for disc accretion} \label{SEC:tSS}%

We model disc accretion using the \citet{Shakura1973} prescription, which conflates all possible angular momentum transport mechanisms into a single parameter, $\alpha_{_{\rm SS}}$. For a low-mass disc in approximate Keplerian rotation around a star of mass $M_{_\star}$, the accretion timescale at radius $R$ is then $\;\sim\alpha_{_{\rm SS}}^{-1}(GM_\star R)^{1/2}a^{-2}$, where $a$ is the local sound speed. We therefore compute a kernel-weighted mean over all the SPH particles in the interaction-zone, 
\begin{eqnarray}\label{EQN:tDISC}
\left<t_{_{\rm DISC}}\right>\!&\!=\!&\!\frac{(GM_s)^{1/2}}{\alpha_{_{\rm SS}}{\cal W}}\!\sum\limits_{j}\!\left\{\!\frac{|\Delta{\bf r}_{js}|^{1/2}m_jW(|\Delta{\bf r}_{js}|,H_s)}{\rho_ja_j^2}\!\right\}.
\end{eqnarray}
{\refrpt Both observational estimates of protostellar accretion disc lifetimes \citep[e.g.][]{Hartmann1998}, and theoretical simulations \citep[e.g.][]{Forgan2010}, suggest that $\alpha_{_{\rm SS}}=0.01$ is a suitable default value, but there are large uncertainties.}

\subsubsection{Net timescale for accretion} \label{SEC:ACCTIMESCALE}%

The timescale for accretion onto the point-mass is given by
\begin{eqnarray}\label{EQN:t_ACC}
t_{_{\rm ACC}}&=&\left<t_{_{\rm RAD}}\right>_s^{(1-f)}\,\left<t_{_{\rm DISC}}\right>_s^f\,,\\
f&=&\mbox{\sc min}\left\{\frac{2E_{_{\rm ROT}}}{\left|E_{_{\rm GRAV}}\right|}\;,\,1\right\}\,.
\end{eqnarray}
Here $E_{_{\rm ROT}}$ and $E_{_{\rm GRAV}}$ are -- respectively -- the net rotational and gravitational energies of the SPH particles in the interaction-zone, relative to the point-mass, i.e.
\begin{eqnarray}
E_{_{\rm ROT}}\!&\!=\!&\!\frac{|{\bf L}_{_{\rm INT}}|^4}{2\sum_j\left\{m_j|\Delta{\bf r}_{js}\!\cdot\!{\bf L}_{_{\rm INT}}|^2\right\}}\,,\\\nonumber
E_{_{\rm GRAV}}\!&\!=\!&\!\frac{GM_s}{2}\!\sum_j\!m_j\!\left\{\!\phi\!\left(\frac{|\Delta{\bf r}_{js}|}{H_s}\!\right)\!+\!\phi\!\left(\!\frac{|\Delta{\bf r}_{js}|}{h_j}\!\right)\!\right\}\\\nonumber
&&\!+\,\frac{G}{4}\!\sum_j\!\sum_{j'\neq j}\!m_jm_{j'}\!\left\{\!\phi\!\left(\!\frac{\left|\Delta{\bf r}_{jj'}\right|}{h_j}\!\right)\!+\!\phi\!\left(\!\frac{\left|\Delta{\bf r}_{jj'}\right|}{h_{j'}}\!\right)\!\right\}\!.\\
\end{eqnarray}
\begin{eqnarray}\label{EQN:LINT}
{\bf L}_{_{\rm INT}}&=&\sum_{j}\left\{m_j\Delta{\bf r}_{js}\!\times\!\Delta{\bf v}_{js}\right\}
\end{eqnarray}
is the net angular momentum of the SPH particles in the interaction-zone, relative to the point-mass, and $\phi$ is a function that encapsulates the kernel-smoothing of the gravitational potential \citep[see Eqn. 15 in][]{Hubber2011}.

{\refrpt Eqn. \ref{EQN:t_ACC} is adopted because it gives the correct limiting behaviour.} If the SPH particles inside the sink are in rotational equilibrium, $t_{_{\rm ACC}}\rightarrow\left<t_{_{\rm DISC}}\right>_s$. Conversely, if they are not rotating at all, $t_{_{\rm ACC}}\rightarrow\left<t_{_{\rm RAD}}\right>_s$.

\subsubsection{Excising SPH particles} \label{SEC:ACCTIMESCALE}%

Normally, the mass accreted by the point-mass at the end of the current timestep, $(t,t+\delta t_s)$, is
\begin{equation} \label{EQN:MACC}
\delta M_{_{\rm ACC}} = M_{_{\rm INT}}\,\left[ 1 - \exp{\left( - \frac{\delta t_s}{t_{_{\rm ACC}} }\right)} \right]\,.
\end{equation}
In the first instance, this mass is removed from the SPH particle closest to the point-mass. If the mass of this SPH particle is less than $\delta M_{_{\rm ACC}}$, the remainder is removed from the second closest SPH particle, and so on, until the whole of $\delta M_{_{\rm ACC}}$ has been removed; {\refrpt because this only affects particles close to the point-mass, it is extremely rare that particles with reduced mass leave the interaction-zone. There are two circumstances under which the procedure outlined above is superseded.}

(i) If the total mass of SPH particles in the interaction-zone of a \NSg presently exceeds the total mass of SPH particles inside the interaction-zone at the time it was created, $M_{_{\rm MAX}}$, then $t_{_{\rm ACC}}$ is decreased artificially by a factor $t_{_{\rm ACC}}\rightarrow t_{_{\rm ACC}}/(M_{_{\rm INT}}/M_{_{\rm MAX}})^2$, in order to accrete the excess mass more rapidly. This is similar to the procedure used to transfer mass from nearby grid cells to sink particles in AMR \citep{Krumholz2004}.

(ii) If the timestep, $\delta t_j$, for an SPH particle, $j$, in the interaction-zone of \NSg $\,s$, satisfies $\delta t_j<\gamma_s\,(R_s^3/GM_s)^{1/2}$, $j$ is immediately accreted -- in its totality -- by $s$. $\gamma_s$ is a tolerance parameter, with default value $0.01$. This device is intended to moderate the situation where the SPH particles inside the interaction-zone have formed a Toomre unstable disc. The assumption is that this will lead to efficient transport of angular momentum by gravitational torques, and hence very rapid inspiral onto the point-mass.

This piecemeal leaching of mass, from the SPH particles in the interaction-zone, onto the point-mass at its centre, is termed regulated accretion, and is a critical feature of the \NSg algorithm. It ensures that, as long as the sink is accreting, there are SPH particles in the interaction-zone, and therefore there are no very steep gradients across the boundary of the sink.

{\refrpt The idea of transferring mass piecemeal from an SPH particle to a sink is not new. This device has been implemented by \citet{Anzer1987}, in simulating the accretion of a supersonic wind by a neutron star, and has subsequently been used in the same context by, for example, \citet{Boffin1994}. In their algorithm, mass is transferred to a pre-existing sink, from all the particles in the computational domain -- but at a rate that is strongly weighted towards those presently near the sink; the weighting function has to be tuned to produce an acceptable rate of accretion onto the sink. However, here we are considering dynamically created sinks and subsonic ambient flows. Moreover, in our algorithm, the particles that leach mass to the point-mass are always the ones near the centre of the interaction-zone, which never leave the interaction-zone; consequently, we only have to contend with a small number of particles having time-varying mass, and none outside of the interaction-zone.}

\subsection{Updating the properties of a \NS}\label{SEC:NEWUPDATE}%

At the end of each timestep, $\delta t_s$, the mass, $M_s$, position, ${\bf r}_s$, velocity, ${\bf v}_s$, and angular momentum, ${\bf L}_s$, of the point-mass are updated to take account of accretion,
\begin{eqnarray}
M_s'&=&M_s\,+\,\sum\limits_{j}\left\{\delta m_j\right\}\,,\\
{\bf r}_s'&=&M_s'^{\;-1}\,\left(M_s{\bf r}_s+\sum\limits_{j}\left\{\delta m_j\,{\bf r}_j\right\}\right)\,,\\
{\bf v}_s'&=&M_s'^{\;-1}\,\left(M_s{\bf v}_s+\sum\limits_{j}\left\{\delta m_j\,{\bf v}_j\right\}\right)\,,\\\nonumber
{\bf L}_s'&=&{\bf L}_s\,+\,M_s\,\Delta{\bf r}_{ss'}\!\times\!\Delta{\bf v}_{ss'}\\\label{EQN:ANGMOM:1}
 &&\hspace{1.0cm}+\,\sum\limits_{j}\left\{\delta m_j\,\Delta{\bf r}_{js'}\!\times\!\Delta{\bf v}_{js'}\right\}\,.
\end{eqnarray}
Here, the summation is over all SPH particles, $j$, that lose mass, $\delta m_j$, to the point-mass, and the updated values are denoted by primes. Once an SPH particle has zero mass, it is removed from the simulation altogether.

\subsection{Angular momentum feedback from a \NS} \label{SEC:NEWANGMOM}%

If the point-mass were to assimilate all the angular momentum of the SPH particles it accreted, as implicit in Eqn. \ref{EQN:ANGMOM:1}, it would spin so fast that it could not reach stellar densities. This is what happens with all standard sinks in SPH: they are sinks of both mass and angular momentum, and this is highly unrealistic. In reality, if the material inflowing towards a protostar has high specific angular momentum, it first falls onto an accretion disc, and then spirals inwards by transferring the bulk of its angular momentum to other material further out in the disc or envelope. Therefore, at the end of each timestep, we reduce the angular momentum of the point-mass by transferring some of its angular momentum to the SPH particles in the surrounding interaction-zone. Since this transfer of angular momentum is presumed to be effected by viscous torques in an accretion disc, the amount of angular momentum transferred is given by
\begin{eqnarray}\label{EQN:ANGMOM:2}
|\delta{\bf L}_s|&=&|{\bf L}_s|\left\{1-\exp\left(-\,\frac{\delta t_s}{\left<t_{_{\rm DISC}}\right>}\right)\right\}\,.
\end{eqnarray}
We achieve this by giving each SPH particle, $j$, in the interaction-zone of $s$ an impulse of velocity
\begin{eqnarray}
\delta{\bf v}_j&=&\frac{|\delta{\bf L}_s|\;{\bf L}_s\!\times\!\Delta{\bf r}_{js}}{\left|\sum_j\left\{m_j\,\Delta{\bf r}_{js}\!\times\!{\bf L}_s\!\times\!\Delta{\bf r}_{js}\right\}\right|}\,.
\end{eqnarray}
Thus each SPH particle in the interaction-zone of $s$ receives an impulse of velocity proportional to its distance from the rotation axis of the point-mass.

In order to compensate for these impulses of velocity, the point-mass receives impulses of momentum and angular momentum given by
\begin{eqnarray}
\delta{\bf v}_s&=&-\;M_s^{-1}\;\sum_j\left\{m_j\,\delta{\bf v}_j\right\}\,,\\
\delta{\bf L}_s&=&-\;\sum_j\left\{m_j\;\Delta{\bf r}_{js}\!\times\!\delta{\bf v}_j\right\}\,.
\end{eqnarray}
As a consequence, the net angular momentum (invested in SPH particle motions and the spins of point-masses) is accurately conserved. At the same time, the amount of angular momentum invested in the spins of point-masses is small, both because it is continually transferred to the SPH particles in the interaction-zone, and because the close-in SPH particles from which it assimilates mass have normally had to lose a lot of angular momentum to get close-in in the first place. This is a critical feature of the \NSg algorithm. It ensures that \NSsg do not act as sinks for angular momentum, and that the point-mass can only grow in mass at a rate influenced by how fast the close-in SPH particles are able to transfer their angular momentum to other SPH particles further out in the interaction-zone and beyond.

\section{The \OSg algorithm} \label{SEC:OLDSINKS}%

Many variants of standard sink have been used previously in SPH simulations of star formation \citep[e.g.][]{Bate1995,Federrath2010,Wadsley2011}, and it is not practical to present test results for all of them. Here, and in Section \ref{SEC:URSINKS}, we describe the particular representative implementations that we use for comparison tests. The choices have been made with a view to distinguishing the problems associated with sink {\it creation} from those associated with sink {\it evolution}. In this Section we define \OSs, which are {\it created} in the same way as \NSs, and therefore avoid the problems normally associated with the creation of standard sinks. This enables us to isolate the problems associated with the evolution of standard sinks, viz. that they all appear to seriously corrupt the hydrodynamics near the sink boundary (by introducing excessively steep gradients), to act as sinks for angular momentum, and to produce results that are strongly dependent on the user-defined parameters of sink creation and evolution. 

\subsection{Internal structure and advection of an \OS} \label{SEC:OLDSTRUCT}%

An \OSg comprises a central point mass (the {\it point-mass}) and a comoving concentric spherical volume of radius $R_s$ (hereafter the {\it exclusion-zone}). The mass, $M_s$, position, ${\bf r}_s$, velocity, ${\bf v}_s$, and angular momentum, ${\bf L}_s$, of an \OSg refer to its point-mass. The exclusion-zone, as its name implies, generally contains very few SPH particles -- and this is one of the critical features that distinguishes an \OSg from a \NS. Once created, an \OSg is tracked with the same integration routine as an SPH particle.

\subsection{Criteria for creation of an \OS} \label{SEC:OLDCREATE}%

The criteria for creation of an \OSg are the same as for creation of a \NS, viz. that an SPH particle, $i$, has density exceeding $\rho_{_{\rm SINK}}$ (Eqn. \ref{EQN:RHOCRIT}), that the resulting sink does not -- at its creation -- overlap a pre-existing sink (Eqn. \ref{EQN:VRLPCRIT}), that $i$ has lower gravitational potential than all its neighbours (Eqn. \ref{EQN:POTMINCRIT}), and that $i$ satisfy the Hill Sphere criterion (Eqn. \ref{EQN:HILLCRIT}). If an SPH particle $i$ satisfies these four criteria, an \OSg is created with the same mass, $M_s\!=\!m_i$, position, ${\bf r}_s\!=\!{\bf r}_i$, velocity, ${\bf v}_s\!=\!{\bf v}_i$, and angular momentum, ${\bf L}_s\!=\!{\bf 0}$, as $i$. The radius of the exclusion-zone is set to $R_{_s}=X_{_{\rm SINK}}h_i$, where $X_{_{\rm SINK}}$ is a user-defined parameter. With the M4 smoothing kernel, the default value is $X_{_{\rm SINK}}=2$, so that the exclusion-zone is the same as the smoothing volume of $i$.

\subsection{Criteria for accretion by an \OS}\label{SEC:OLDACCRETE}%

An SPH particle $j$ is assimilated by the point-mass of an \OS, $s$, if it satisfies two criteria.

First, it must have entered the exclusion-zone of $s$, i.e.
\begin{eqnarray}\label{EQN:OLDACCRETE:1}
|{\bf r}_j-{\bf r}_s|&<&R_s\,.
\end{eqnarray}

Second, the mutual non-thermal energy of $j$ and $s$ (i.e. their mutual bulk-kinetic plus gravitational energy),
\begin{eqnarray}\nonumber
e_{js}\!\!&\!\!=\!\!\!&\!\frac{m_jM_s|\Delta{\bf v}_{js}|^2}{2(m_j+M_s)}\!+\!\frac{Gm_jM_s}{2}\!\left\{\!\phi\!\left(\!\frac{|\Delta{\bf r}_{js}|}{h_j}\!\right)\!+\!\phi\!\left(\!\frac{|\Delta{\bf r}_{js}|}{H_s}\!\right)\right\}\!,\\\label{EQN:OLDACCRETE:2}
\end{eqnarray}
must be negative. If more than one \OSg is minded to accrete $j$, $j$ is accreted by the one whose point-mass is closest. When an SPH particle $i$ triggers the formation of an \OS, $s$, these conditions are normally satisfied by all, or most, of the neighbours of $i$, so $s$ immediately assimilates all, or most, of the neighbours of $i$.

\citet{Bate1995} use two additional criteria to determine whether an SPH particle, $j$ is assimilated by a standard sink, $s$. First, the angular momentum of $j$ relative to the point-mass of $s$ should be less than it would be if $j$ were in a circular orbit of radius $R_s$ about the point-mass, i.e.
\begin{eqnarray}\label{EQN:ANGMOMCRIT}
\left|\Delta{\bf r}_{js}\!\times\!\Delta{\bf v}_{js}\right|&<&\left(GM_sR_s\right)^{1/2}\,.
\end{eqnarray}
Second, if an SPH particle $j$ passes within a small distance ($0.1R_s$ or less) of the point-mass, all other criteria are ignored, and it is assimilated by the point-mass. We find that these criteria do not have a significant effect on the results.

\citet{Bate1995} have also developed a procedure for extrapolating the density and velocity fields outside a sink, and using this information to generate correction terms to account for the missing SPH particles inside the exclusion-zone. However, these terms have not been included in subsequent implementations of the \citet{Bate1995} algorithm (Bate, private communication), and we also do not include them in \OSsg (or \USs, see Section \ref{SEC:URSINKS}).

\subsection{Updating the properties of an \OS} \label{SEC:OLDUPDATE}%

Accretion of an SPH particle by an \OS, $s$, is immediate and complete; the SPH particle is removed from the simulation, and the properties of $s$ are updated according to
\begin{eqnarray}\label{EQN:OLDMASSUPDATE}
M_s'&=&M_s\,+\,\sum\limits_{j}\left\{m_j\right\}\,,\\
{\bf r}_s'&=&M_s'^{\;-1}\,\left(M_s{\bf r}_s+\sum\limits_{j}\left\{m_j\,{\bf r}_j\right\}\right)\,,\\
{\bf v}_s'&=&M_s'^{\;-1}\,\left(M_s{\bf v}_s+\sum\limits_{j}\left\{m_j\,{\bf v}_j\right\}\right)\,,\\\nonumber
{\bf L}_s'&=&{\bf L}_s\,+\,M_s\,\Delta{\bf r}_{ss'}\!\times\!\Delta{\bf v}_{ss'}\\\label{EQN:OLDANGMOMUPDATE}
 &&\hspace{1cm}+\,\sum\limits_{j}\left\{m_j\,\Delta{\bf r}_{js'}\!\times\!\Delta{\bf v}_{js'}\right\}\,;
\end{eqnarray}
the primed variables represent the new properties of $s$, and the sums are over all the SPH particles $j$ accreted during that timestep. We term this instantaneous accretion.

We note that, if an \OSg grows by disc accretion, most of the SPH particles that it accretes only just satisfy Eqn. \ref{EQN:OLDACCRETE:1}, and so its specific angular momentum is
\begin{eqnarray}\nonumber
\frac{\left|{\bf L}_s\right|}{M_s}&\sim&10^{20}\,{\rm cm}^2\,{\rm s}^{-1}\,\left(\frac{M_s}{{\rm M}_{_\odot}}\right)^{1/2}\\
&&\hspace{0.2cm}\times\,\left(\frac{m_{_{\rm SPH}}}{10^{-5}\,{\rm M}_{_\odot}}\right)^{1/6}\,\left(\frac{\rho_{_{\rm SINK}}}{10^{-13}\,{\rm g}\,{\rm cm}^{-3}}\right)^{-1/6}.
\end{eqnarray}
This is so large that the point-mass can not condense to stellar densities (for example, the Sun spinning at break-up speed only has specific angular momentum $\sim\!(G{\rm M}_{_\odot}{\rm R}_{_\odot})^{1/2}\!\sim\!3\times 10^{18}\,{\rm cm}^2\,{\rm s}^{-1}$), unless $m_{_{\rm SPH}}$ is very small (impractically high mass-resolution) and/or $\rho_{_{\rm SINK}}$ is very high (so high as to negate the advantages of using sinks).

\section{The \USg algorithm}\label{SEC:URSINKS}%

In this Section we define \USs, which are created {\it and} evolved in the manner of standard sinks, and are representative of the sinks that have been used in almost all extant SPH simulations of star formation.

\subsection{Internal structure and advection of an \US}\label{SEC:URSTRUCT}%

An \USg has the same internal structure as an \OS, viz. a central point-mass surrounded by a comoving concentric spherical exclusion-zone of radius $R_s$. The point-mass has mass, $M_s$, position, ${\bf r}_s$, velocity, ${\bf v}_s$ and angular momentum ${\bf L}_s$, and is advected like an SPH particle.

\subsection{Criteria for creation of an \US}\label{SEC:URCREATE}%

Some of the criteria for creation of an \USg are different from those used for \NSsg and \OSs. If an SPH particle $i$ satisfies the five criteria detailed below, an \USg is created with the same mass, $M_s\!=\!m_i$, position, ${\bf r}_s\!=\!{\bf r}_i$, velocity, ${\bf v}_s\!=\!{\bf v}_i$, and angular momentum, ${\bf L}_s\!=\!{\bf 0}$, as $i$. The radius of the exclusion-zone is set to $R_{_s}=X_{_{\rm SINK}}h_i$, and if the M4 smoothing kernel is used, the default value is $X_{_{\rm SINK}}=2$, so that the exclusion-zone is the same as the smoothing volume of $i$.

\subsubsection{Density and overlap criteria}\label{SEC:URDENSCRIT}%

The first two criteria for SPH particle $i$ to trigger the creation of an \USg are the same as for \NSsg and \OSs, viz. that its density  should exceed a user-defined threshold, and that the resulting \USg should not overlap a pre-existing \USg (see Sections \ref{SEC:RHOCRIT} and \ref{SEC:VRLPCRIT}).

\subsubsection{$\nabla\!\cdot\!{\bf a}$ criterion}\label{SEC:URDIVACCCRIT}%

The third criterion for SPH particle $i$ to trigger the creation of an \USg is that the divergence of the acceleration at the position of $i$ should be negative,
\begin{eqnarray}\label{EQN:DIVACCCRIT}
(\nabla\!\cdot\!{\bf a})_i&<&0\,,
\end{eqnarray}
otherwise the possibility exists that the dense gas is about to be torn apart tidally. This criterion is used by \citet{Bate1995} and \citet{Wadsley2011}.

\subsubsection{$\nabla\!\cdot\!{\bf v}$ criterion}\label{SEC:URDIVVELCRIT}%

The fourth criterion for SPH particle $i$ to trigger the creation of an \USg is that the divergence of the velocity at the position of $i$ should be negative,
\begin{eqnarray}\label{EQN:DIVVELCRIT}
(\nabla\!\cdot\!{\bf v})_i&<&0\,,
\end{eqnarray}
otherwise the density there must be decreasing. This criterion is used by \citet{Wadsley2011}.

\subsubsection{Non-thermal energy criterion}\label{SEC:URENERGYCRIT}%

The fifth and final criterion for SPH particle $i$ to trigger the creation of an \USg is that the net non-thermal energy of $i$ and its neighbours $j$, in the centre-of-mass frame, should be negative,
\begin{eqnarray}\label{EQN:ENERGYCRIT}
E_{_{\rm NT}}&=&E_{_{\rm GRAV}}+E_{_{\rm KIN}}\;\,<\;\,0\,.
\end{eqnarray}
Here,
\begin{eqnarray}\nonumber
E_{_{\rm GRAV}}\!&\!=\!&\!\frac{G}{4}\!\sum_j\!\sum_{j'\neq j}\!m_jm_{j'}\!\left\{\!\phi\!\left(\!\frac{\left|\Delta{\bf r}_{jj'}\right|}{h_j}\!\right)\!+\!\phi\!\left(\!\frac{\left|\Delta{\bf r}_{jj'}\right|}{h_{j'}}\!\right)\!\right\}\\
\end{eqnarray}
is the self-gravitational potential energy,
\begin{eqnarray}
E_{_{\rm KIN}}&=&\frac{1}{2}\,\sum\limits_{j}\left\{m_j|{\bf v}_j-\overline{\bf v}_j|^2\right\}
\end{eqnarray}
is the net kinetic energy, and $\overline{\bf v}_j\!=\!\sum\left\{m_j{\bf v}_j\right\}/\sum\left\{m_j\right\}$ is the centre-of-mass velocity. Here all the sums over the neighbours $j$ include $i$ itself. This criterion is somewhat weaker than the energy criteria used by other standard algorithms (see Section \ref{SEC:ALTCREATECRIT} below).

\subsubsection{Alternative creation criteria}\label{SEC:ALTCREATECRIT}%

In this Section, we list -- for completeness -- some of the additional criteria used in other standard sink creation algorithms, but not for \USs.

\citet{Bate1995} and \citet{Wadsley2011} apply additional energy criteria for creating a standard sink, viz.
\begin{eqnarray}
E_{_{\rm THERM}}+\frac{1}{2}\,E_{_{\rm GRAV}}&<&0\,,\label{EQN:ENERGYCOND:1} \\
E_{_{\rm ROT}}+E_{_{\rm THERM}}+E_{_{\rm GRAV}}&<&0\,,\label{EQN:ENERGYCOND:2} \\
E_{_{\rm KIN}}+E_{_{\rm THERM}}+E_{_{\rm GRAV}}&<&0\,.\label{EQN:ENERGYCOND:3}
\end{eqnarray}
Here
\begin{eqnarray}
E_{_{\rm THERM}}&=&\sum\limits_{j}\left\{\frac{m_j\,a_j^2}{(\gamma_j-1)}\right\}
\end{eqnarray}
is the net thermal energy ($a_j$ and $\gamma_j$ are the isothermal sound speed and ratio of specific heats for SPH particle $j$),
\begin{eqnarray}
E_{_{\rm ROT}}&=&\frac{|{\bf L}_{_{\rm INT}}|^4}{2\sum_j\left\{m_j|\Delta{\bf r}_{js}.{\bf L}_{_{\rm INT}}|^2\right\}}
\end{eqnarray}
is the net rotational energy,
\begin{eqnarray}
{\bf L}_{_{\rm INT}}&=&\sum_{j}\left\{m_j({\bf r}_j-\overline{\bf r}_j)\!\times\!({\bf v}_j-\overline{\bf v}_j)\right\}
\end{eqnarray}
is the net angular momentum, and $\overline{\bf r}_j\!=\!\sum\left\{m_j{\bf r}_j\right\}/\sum\left\{m_j\right\}$ is the centre of mass. All the sums over the neighbours $j$ include $i$ itself. Although these energy criteria are somewhat stiffer than Eqn. (\ref{EQN:ENERGYCRIT}), they do not produce significantly different reaults.

\citet{Federrath2010} and \cite{Wadsley2011} also use the potential-minimum criterion (i.e. Eqn. \ref{EQN:POTMINCRIT}), and in addition they advocate checking that the flow is convergent in all directions, by computing the eigenvalues of $dv_i/dx_j$ (here $i$ and $j$ are identifiers for the Cartesian coordinates) and requiring them to be individually negative. However, these criteria have not been implemented in the majority of extant SPH simulations of star formation.

\subsection{Criteria for accretion by an \US}\label{SEC:URACCRETE}%

An SPH particle $j$ is immediately accreted by the point-mass of an \USg $s$, if it falls within the exclusion-zone of $s$, {\it and} their mutual non-thermal energy is negative, just as for an \OSg (Eqns. \ref{EQN:OLDACCRETE:1} \& \ref{EQN:OLDACCRETE:2}). Under normal circumstances, a newly-created \USg immediately assimilates most of the neighbours of the SPH particle that triggered its creation.

\subsection{Updating the properties of an \US}\label{SEC:URUPDATE}%

Each time an \USg assimilates one or more SPH particles, the properties of the \USg are updated according to Eqns. (\ref{EQN:OLDMASSUPDATE}) through (\ref{EQN:OLDANGMOMUPDATE}), just as for an \OS. As with an \OS, this often leads to the unphysical acquisition of large amounts of angular momentum.

\begin{table*}
\begin{center}
\begin{tabular}{r|c|ccccc}\hline
{\sc Test}$\;\longrightarrow$                 & \hspace{0.3cm} & \hspace{0.3cm}Default\hspace{0.3cm} & 
$\;\;$Bondi Accretion$\;\;$                               & Rotating Bonnor-Ebert & Boss-Bodenheimer                 & Turbulent Core \\\hline
$\rho_{_{\rm SINK}}/({\rm g}\,{\rm cm}^{-3})$ &                & $10^{-11}$                          & 
\dsh                                                      & $10^{-11}$            & $10^{-13},\,10^{-12},\,10^{-11}$ & $10^{-11},\,10^{-10},\,10^{-9}$      \\
$X_{_{\rm SINK}}$                             &                & $2$                                 & 
\dsh                                                      & $2$                   & $2,\,4$                          & $2$            \\
$X_{_{\rm HILL}}$                             &                & $4$                                 & 
\dsh                                                      & $4$                   & $4$                              & $4$            \\
$\alpha_{_{\rm SS}}$                          &                & $0.01$                              & 
$0.01$                                                    & $0.01$                & $0.01$                           & $0.01$         \\
$R_s/R_{_{\rm SONIC}}$                        &                & \dsh                                & 
$\frac{1}{8},\,\frac{1}{4},\,\frac{1}{2},\,1,\,2,\,4,\,8$ & \dsh                  & \dsh                             & \dsh           \\\hline
\NS                                           &                &                                     &
$\bigstar$                                                & $\bigstar$            & $\bigstar$                       & $\bigstar$     \\
\OS                                           &                &                                     &
$\bigstar$                                                & $\bigstar$            & $\bigstar$                       & $\bigstar$     \\
\US                                           &                &                                     &
($\bigstar$)                                              & ($\bigstar$)          & ($\bigstar$)                     & $\bigstar$     \\\hline
\end{tabular}
\caption{Default parameter values and values used for the different tests. Row 1 gives the name of the test. Rows 2 to 6 give the parameter values, and rows 7 to 9 indicate which types of sink were tested, with brackets indicating the tests that are not discussed.}
\end{center}
\label{TAB:PARAMS}
\end{table*}

\section{Tests}\label{SEC:TESTS}%

We have used the following four tests to compare \NSsg with \OSsg and \USs, and to explore the effect of changing the user-prescribed parameters for sink creation ($\rho_{_{\rm SINK}}$ and $X_{_{\rm SINK}}$) and the SPH resolution (${\cal N}_{_{\rm SPH}}$): isothermal Bondi accretion (Section \ref{SEC:BONDI}), the collapse of a rotating Bonnor-Ebert sphere (\ref{SEC:BONNOREB}), the Boss-Bodenheimer test (\ref{SEC:BOSSBOD}), and the fragmentation of a turbulent prestellar core (\ref{SEC:TURBCORE}). Parameter values used for the different tests are summarised in Table 2. Although all four tests have been performed with all three types of sink, we do not present and discuss results obtained with \USsg for the first three tests; this is because the results obtained with \USsg are very chaotic and show no sign of converging; the turbulent prestellar core test suffices to demonstrate this.

All the tests are performed with the {\sc seren} SPH code, using its default options \citep{Hubber2011}. We set $\eta\!=\!1.2$, so that the smoothing-length of SPH particle $i$ is $h_i\!=\!1.2(m_{_{\rm SPH}}/\rho_i)^{1/3}$ and the mean number of neighbours is $\bar{\cal N}_{_{\rm NEIB}}\sim\! 58$.

\subsection{Bondi accretion} \label{SEC:BONDI}%

The first test models Bondi accretion. The accretion rate for the transsonic isothermal solution \citep[][]{Bondi1952} is
\begin{eqnarray}
\dot{M}_{_{\rm BO}}&=&\frac{{\rm e}^{3/2}\,\pi\,G^2\,M_\star^2\,\rho_{_{\rm O}}}{a_{_{\rm O}}^3}\,,
\end{eqnarray}
where {\refrpt e is the base of natural logarithms,} $M_\star$ is the mass of the central star, $\rho_{_{\rm O}}$ is the density at infinity, and $a_{_{\rm O}}$ is the isothermal sound speed. The solution is obtained by neglecting the self-gravity of the inflowing gas. A critical role is played by the sonic radius at 
\begin{eqnarray}
R_{_{\rm SONIC}}&=&\frac{GM_\star}{2a_{_{\rm O}}^2}\,.
\end{eqnarray}
For $r \gg R_{_{\rm SONIC}}$ the inward flow is subsonic and the pressure acceleration ($-\rho^{-1}\nabla P$) plays an important role. Conversely, for $r \ll R_{_{\rm SONIC}}$ the inward flow approaches free fall and the pressure acceleration is unimportant. 

\begin{figure} 
\begin{center}
\includegraphics[width=85mm,angle=0]{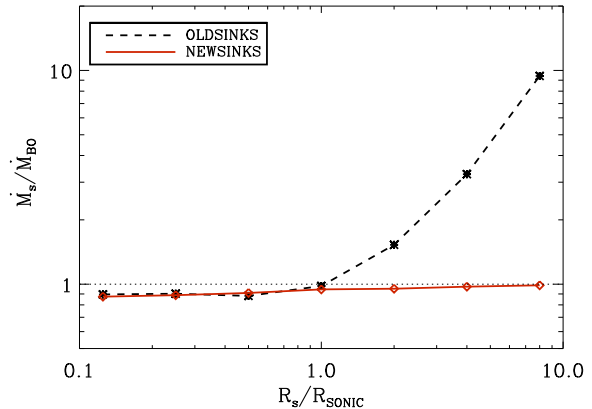}
\caption{The normalised accretion rate ($\dot{M_s}/\dot{M}_{\rm BO}$) as a function of the normalised sink radius ($R_s/R_{_{\rm SONIC}}$) for isothermal Bondi accretion. Results obtained with \NSsg are shown as red diamonds connected by a solid line, and those obtained with \OSsg are shown as black stars connected by a dashed line.}
\label{FIG:BONDI}
\end{center}
\end{figure}

The density and velocity profiles for the transsonic solution are obtained by solving the Bernouilli equation numerically. The initial conditions are set up by relaxing a periodic cube to produce a uniform-density glass, cutting a sphere containing $5 \times 10^5$ particles from this cube, stretching the sphere to produce the required density field, and giving each SPH particle the inward radial velocity appropriate to its position. The outer boundary of the sphere is at $\sim\!20\,R_{_{\rm SONIC}}$, and the simulations are followed to time $t_{_{\rm END}}=2GM_\star/a_{_{\rm O}}^3$.

This test does not involve sink creation, and therefore we do not specify $\rho_{_{\rm SINK}}$ or $X_{_{\rm SINK}}$. Instead, a sink is placed at the origin from the outset. Its point-mass has mass $M_s\!=\!{\rm M}_{_\odot}$ and, although it accretes SPH particles and has an accretion rate, $\dot{M}_s$, its mass is held fixed. This is self-consistent, since, if the accreting gas is sufficiently rarefied to have negligible self-gravity (compared with the gravity of the central star), the increase in the central mass by $t_{_{\rm END}}$ is negligible. {\refrpt The purpose of this test is to evaluate the treatment of radial accretion.} 

The system is evolved with a \NSg or an \OS, with the sink radius set to different multiples of the sonic radius, $R_s/R_{_{\rm SONIC}}=1/8,\,1/4,\,1/2,\,1,\,2,\,4,\,8$. By $t_{_{\rm END}}$, the accretion rate is steady, and the rarefaction wave propagating in from the outer boundary is still in the outer parts of the computational domain. Fig. \ref{FIG:BONDI} shows how the accretion rates depend on $R_s/R_{_{\rm SONIC}}$. Since this test does not involve sink creation, the results obtained with an \USg are exactly the same as those obtained with an \OS.

As long as $R_s<R_{_{\rm SONIC}}$, both the \NSg and the \OSg model Bondi accretion well, giving accretion rates $\sim\!10\%$ below the analytic value. Under this circumstance, the material flowing into the sink is close to free fall and so the pressure acceleration is unimportant; the fact that \OSsg give rise to inaccurate evaluations of the pressure and viscous forces near $R_s$ is then of little consequence.

However, for $R_s>R_{_{\rm SONIC}}$, an \OSg seriously overestimates the accretion rate. Under this circumstance, the pressure acceleration is an important factor controlling the flow of material into the sink, and the steep outward pressure gradient at the edge of an \OSg artificially increases the accretion rate; for $R_s/R_{_{\rm SONIC}}\!=\!8$, $\dot{M}_s$ is ten times the analytic value. In contrast, with a \NSg there is a small increase in the accretion rate with increasing $R_s$, but, even for $R_s/R_{_{\rm SONIC}}=8$, $\,\dot{M}_s$ is within $\sim\!1\%$ of the analytic value.

Although this test only deals with one possible accretion mode, all spherical accretion modes are likely to suffer from the same problem, if the inflow is subsonic near the sink boundary. This is a serious concern in simulations of star formation that set $\rho_{_{\rm SINK}}$ high, in order to follow protostars well into their Kelvin-Helmholtz contraction phase. The flow of gas into the sink is then subsonic, and an \OSg will experience artificially enhanced accretion. A \NSg should capture the accretion rate much more accurately.

\begin{figure*}
\begin{center}
\includegraphics[width=180mm]{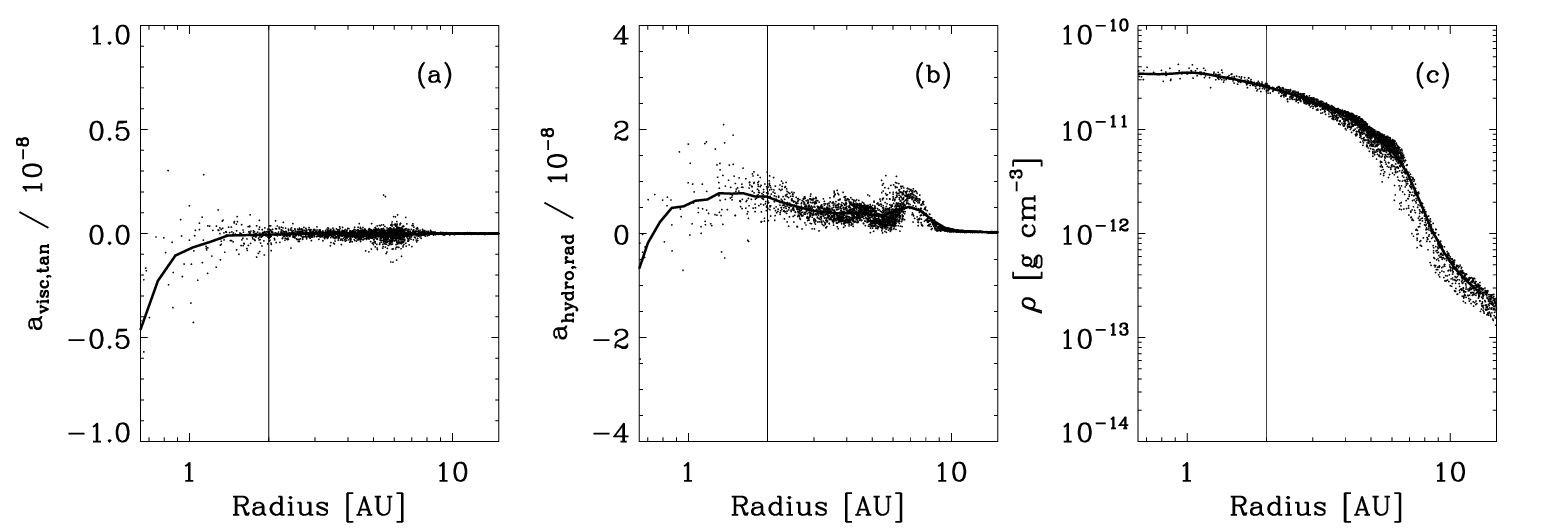}\\
\includegraphics[width=180mm]{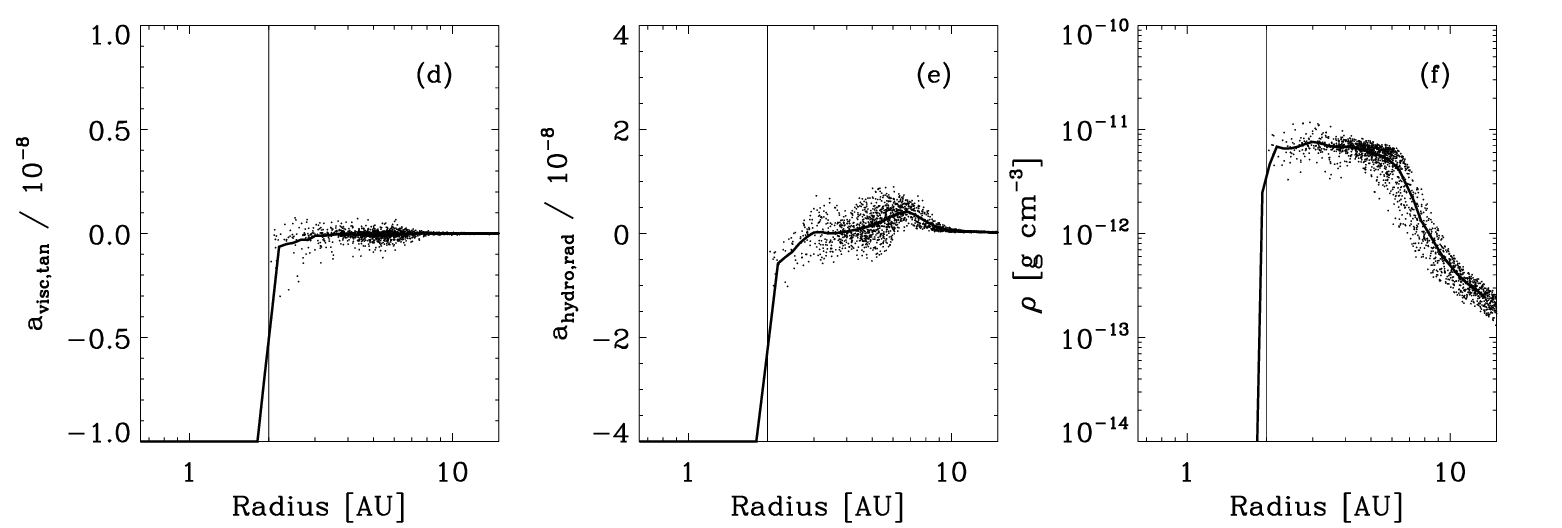}\\
\caption{(a,d) Tangential viscous accelerations, (b,e) radial hydrodynamic accelerations, and (c,f) densities, for all the SPH particles within $\sim\!20\,{\rm AU}$ of the point-mass, shortly after sink creation in the Rotating Bonnor-Ebert Sphere Test. The results on the top row have been obtained with a \NS, and those on the bottom row with an \OS.}
\label{FIG:BONNOREB:1}
\end{center}
\end{figure*}

\begin{figure*}
\begin{center}
 \includegraphics[width=80mm]{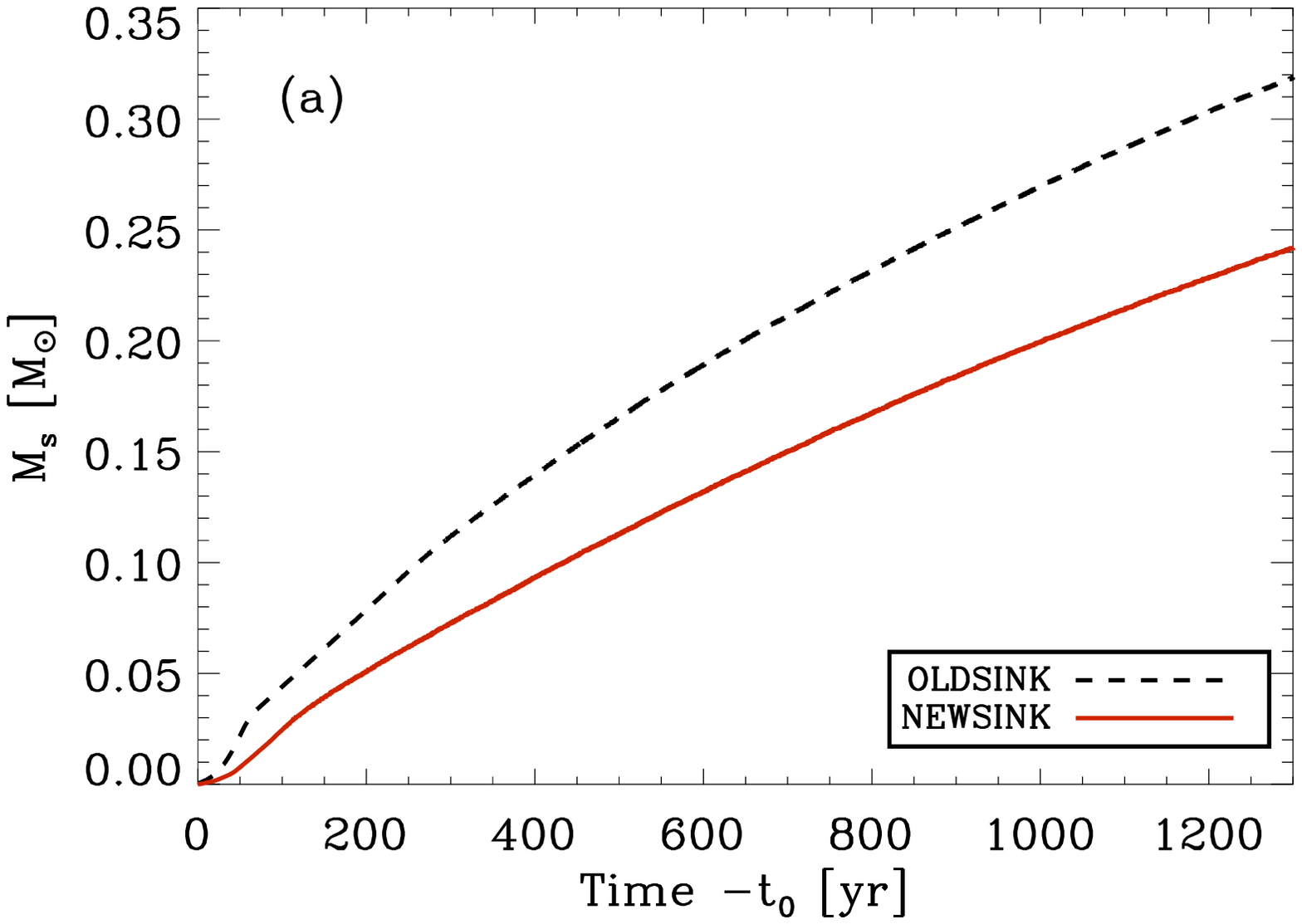} 
 \includegraphics[width=80mm]{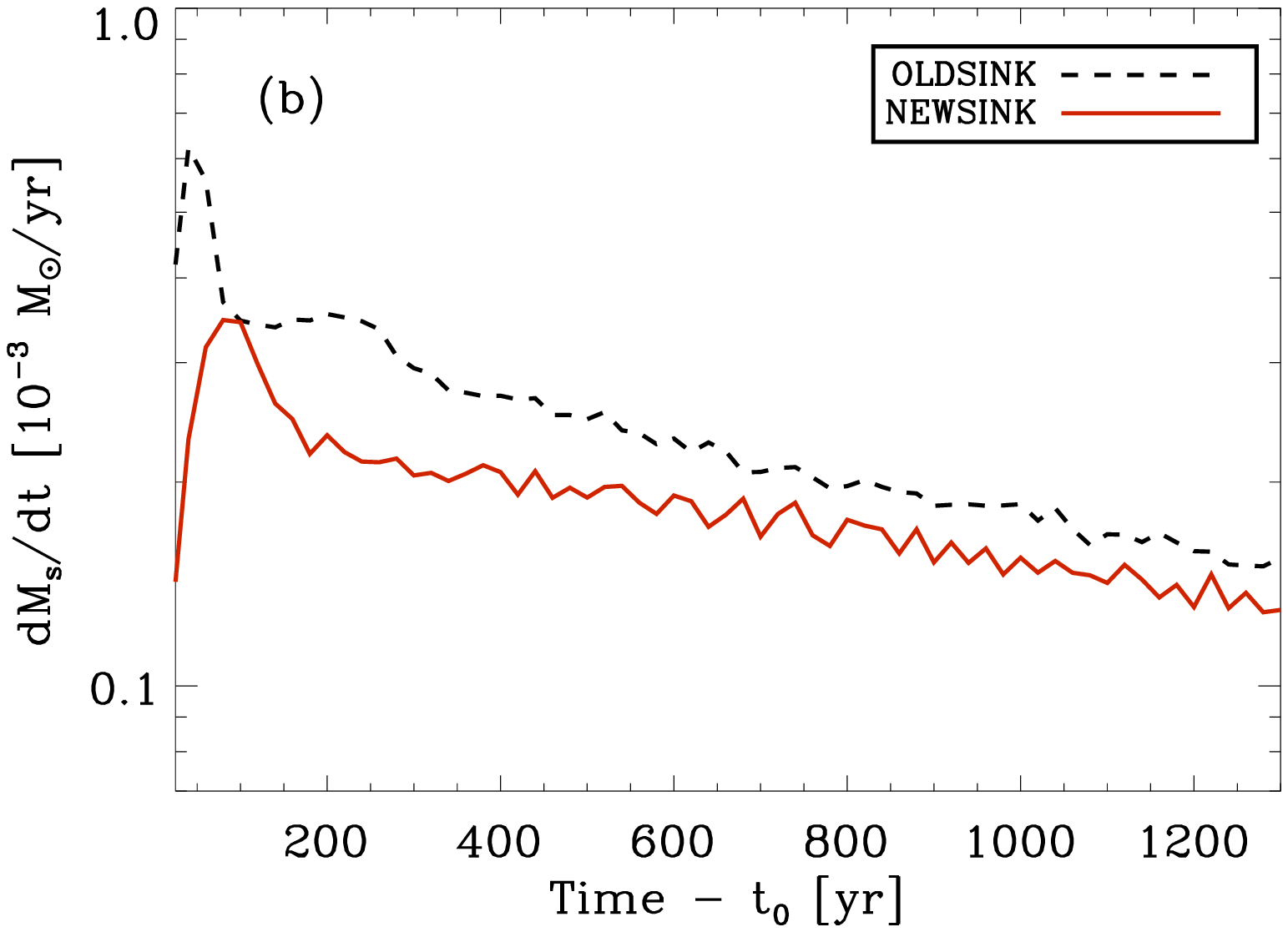} 
\caption{(a) The sink mass, and (b) the sink accretion rate, in the Rotating Bonnor-Ebert Sphere Test, as a function of time elapsed since the formation of the sink at $t_0$, using a \NSg (red solid lines) and an \OSg (black dashed lines).}
\label{FIG:BONNOREB:2}
\end{center}
\end{figure*}

\subsection{Rotating Bonnor-Ebert sphere} \label{SEC:BONNOREB}%

The main mode of accretion in star formation involves discs. The accretion rate is then controlled by the torques that redistribute angular momentum. There are no analytic solutions for this mode, so testing is less straightforward. The test we perform involves a rigidly rotating spherical gas cloud having mass ${\rm M}_{_\odot}$, initial boundary radius $10^3\,{\rm AU}$, and initial angular speed $3.54\times 10^{-12}\,{\rm s}^{-1}$. The gas is {\refrpt and remains} isothermal with sound speed $0.20\,{\rm km}\,{\rm s}^{-1}$ (corresponding to molecular gas at $T\!=\!11\,{\rm K}$). {\refrpt However, the main purpose of this test is not to explore sink creation from isothermal gas, but rather to evaluate the treatment of disc accretion.}

The initial density profile is that of a critical Bonnor-Ebert Sphere, i.e. one with dimensionless boundary radius $\xi_{_{\rm B}}\!=\!6.45$, but the cloud is not in equilibrium. The initial ratios of thermal and rotational energy to gravitational energy are $\alpha\!=\!0.09$ and $\beta\!=\!0.59$. The cloud is modelled with $10^5$ SPH particles (hence $m_{_{\rm SPH}}\!=\!10^{-5}\,{\rm M}_{_\odot}$), and we set $\rho_{_{\rm SINK}}\!=\!10^{-11}\,{\rm g}\,{\rm cm}^{-3}$ and $X_{_{\rm SINK}}\!=\!2$. The initial density profile is obtained by integrating the Isothermal Equation \citep[e.g.][]{Chandrasekhar+Wares1949} from $\xi\!=\!0$ to $\xi\!=\!6.45$, and scaling the result to give total mass ${\rm M}_{_\odot}$ and radius $10^3\,{\rm AU}$. The initial conditions are then generated by relaxing a periodic cube of SPH particles to produce a uniform glass, cutting a sphere containing $10^5$ particles from this cube, and stretching the particle positions to produce this density profile. The particles start from rest.

The cloud collapses to form a single sink surrounded by a small accretion disc. Fig. \ref{FIG:BONNOREB:1} shows, for all the SPH particles within $\sim\!20\,{\rm AU}$ of the point-mass, the tangential viscous acceleration, $a_{_{\rm TV}}$, the radial hydrodynamic acceleration, $a_{_{\rm RH}}$, and the density, $\rho$, all as functions of distance from the point-mass, shortly after sink creation.  The upper row gives results obtained with a \NS, showing that the SPH particles near $R_s\;(=\!2\,{\rm AU})$ have smoothly varying $a_{_{\rm TV}}$, $a_{_{\rm RH}}$ and $\rho$, due to the presence of SPH particles in the interaction-zone; in particular, $a_{_{\rm RH}}$ is positive (i.e. outward, therefore resisting accretion, as it should). The lower row gives results obtained with an \OS, showing that SPH particles near the sink boundary experience artificially reduced $a_{_{\rm TV}}$, $a_{_{\rm RH}}$ and $\rho$, due to a lack of neighbours in the exclusion-zone; all three factors act to increase the net inward acceleration, and thereby to artificially increase the rate of accretion. In addition, because the SPH particles accreted by an \OSg carry their angular momentum with them, rather than transferring it to the SPH particles a bit further out, these SPH particles can more rapidly move inward, to be accreted in their turn. When this test is performed with \USs, the results are chaotic, and involve the creation of many different sinks; we do not discuss these results.

Fig. \ref{FIG:BONNOREB:2} shows the evolution of the sink mass, $M_s$, and the sink accretion rate, $\dot{M}_s$, as functions of the time elapsed since sink formation, showing that the artificially increased accretion obtained with an \OSg is not a transient effect. By the end of the simulation, which corresponds to $\sim\!100$ orbital periods at the sink boundary, the \OSg has $M_s\!\simeq\!0.37\,{\rm M}_{_\odot}$, whereas the \NSg only has $M_s\!\simeq\!0.24\,{\rm M}_{_\odot}$. Thus a numerical simulation using \OSsg overestimates accretion rates. The effect on final protostellar masses is hard to call. An enhanced accretion rate would mean enhanced radiative and mechanical feedback (if these were included in the simulation), and this might actually terminate accretion earlier, leading to a lower final mass.

\begin{figure*}
\begin{center}
\includegraphics[width=175mm,angle=0]{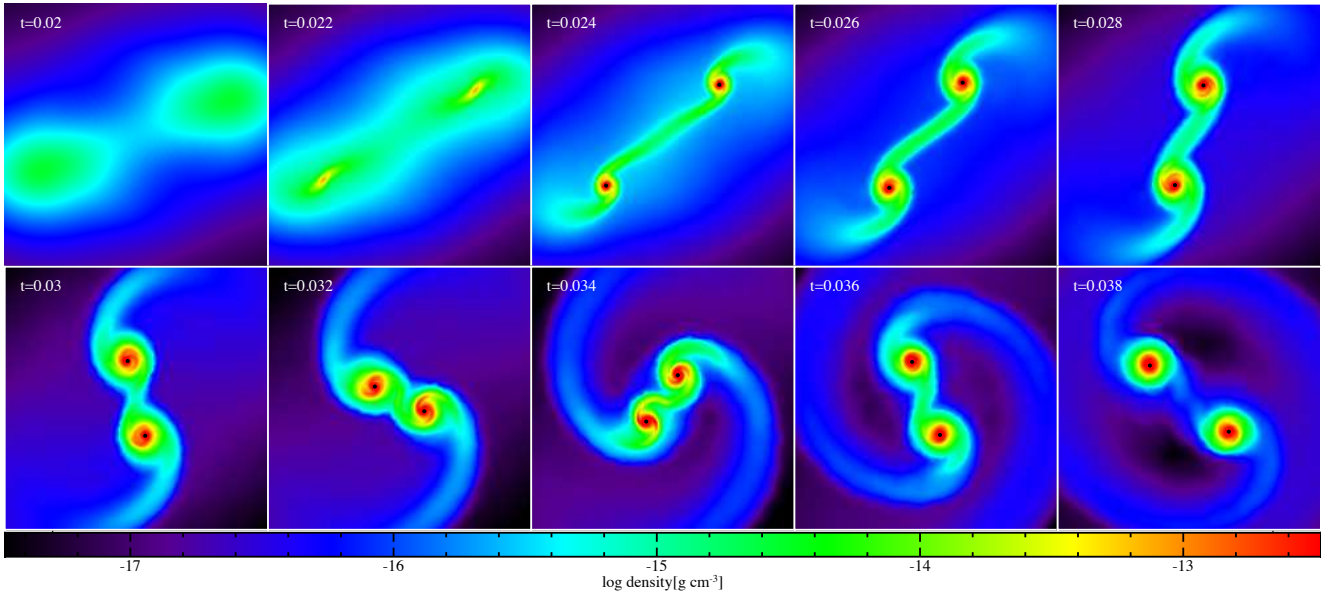}
\caption{A time-sequence of false-colour images of the Boss-Bodenheimer test, performed with \NSs, $X_{_{\rm SINK}}\!=\!2$, $\rho_{_{\rm SINK}}\!=\!10^{-11}\,{\rm g}\,{\rm cm}^{-3}$ and ${\cal N}_{_{\rm SPH}}\!=\!102400$. Each frame has dimensions $(2.4\times 10^3\,{\rm AU})^2$, and the time (in Myrs) is given in the top-left corner. The colour encodes $\log_{_{10}}(\bar{\rho}/{\rm g}\,{\rm cm}^{-3})$, where $\bar{\rho}$ is the density-weighted mean density along the line of sight, and black dots mark sinks.}
\label{FIG:BOSSBOD:1}
\end{center}
\end{figure*}

\subsection{Boss-Bodenheimer test} \label{SEC:BOSSBOD}%

The Boss-Bodenheimer test \citep{BBSIT1979} is a standard test, applied to many star formation codes. For example, \citet{Bate1995} used it to test the original sink algorithm, although of necessity they could only use a rather small number of SPH  particles (${\cal N}_{_{\rm SPH}}\simeq 8\times 10^3$); \citet{BateBurkert1997} repeated the test with ten times as many SPH particles (${\cal N}_{_{\rm SPH}}\simeq 8\times 10^4$), but did not use sinks.

The version we consider here starts with a spherical cloud having mass ${\rm M}_{_\odot}$, radius $2\times 10^3\,{\rm AU}$, and density field
\begin{eqnarray}
\rho(r,\theta,\phi)&=&1.6\times 10^{-17}\,{\rm g}\,{\rm cm}^{-3}\,\left\{1\,+\,0.5\cos(2\phi)\right\}\,,
\end{eqnarray}
where $(r,\theta,\phi)$ are spherical polar coordinates. It is in solid-body rotation with angular speed $\Omega_{_{\rm O}}=1.56\times 10^{-12}\,{\rm s}^{-1}$, and the ratios of thermal and rotational energy to gravitational energy are $\alpha=0.25$ and $\beta=0.20$. The gas obeys a barotropic equation of state of the form
\begin{eqnarray}
T&=&10\,{\rm K}\,\left\{1\,+\,\left(\frac{\rho}{10^{-14}\,{\rm g}\,{\rm cm}^{-3}}\right)^{2/5}\right\}\,,
\end{eqnarray}
corresponding to a molecular gas with ratio of specific heats $7/5$ that is approximately isothermal at low densities and approximately adiabatic at higher densities. The minimum Jeans mass for this equation of state is $M_{_{\rm MIN}}=0.025\,{\rm M}_{_\odot}$.

This test is performed with \NSsg and \OSs, using $X_{_{\rm SINK}}\!=\!2$ and 4 (see Eqn. \ref{EQN:SINKRAD}), $\rho_{_{\rm SINK}}\!=\!10^{-13}$, $10^{-12}$, and $10^{-11}\,{\rm g}\,{\rm cm}^{-3}$, and ${\cal N}_{_{\rm SPH}}\!=\!6400$, 12800, 25600, 51200, 102400, 204800 and 409600. Putting $m_{_{\rm SPH}}={\rm M}_{_\odot}/{\cal N}_{_{\rm SPH}}$ this gives $2.4\times 10^{-6}\,{\rm M}_{_\odot}$ $\la\!m_{_{\rm SPH}}\!\la$ $1.6\times 10^{-4}\,{\rm M}_{_\odot}$. Using
\begin{eqnarray}\label{EQN:SINKRAD:2}
R_s\!&\!=\!&\!X_{_{\rm SINK}}\eta\left(\frac{m_{_{\rm SPH}}}{\rho_{_{\rm SINK}}}\right)^{1/3}\\\label{EQN:SINKRAD:3}
&\!\rightarrow\!&\!1\,{\rm AU}\,X_{_{\rm SINK}}\!\left(\frac{\rho_{_{\rm SINK}}}{10^{-11}\,{\rm g}\,{\rm cm}^{-3}}\right)^{-1/3}\!\left(\frac{{\cal N}_{_{\rm SPH}}}{10^5}\right)^{-1/3}
\end{eqnarray}
we see that $R_s$ ranges from $\sim\!1.3\,{\rm AU}$ (small $X_{_{\rm SINK}}$, large ${\cal N}_{_{\rm SPH}}$, large $\rho_{_{\rm SINK}}$) to $\sim\!46\,{\rm AU}$ (the opposite extremes). {\refrpt Since, in all cases $\rho_{_{\rm SINK}}>10^{-14}\,{\rm g}\,{\rm cm}^{-3}$, sinks are only created in this test once the gas is well into the adiabatic regime.}

The initial conditions are set up by cutting a sphere of $10^5$ SPH particles from a relaxed periodic cube, and scaling the mass and radius to ${\rm M}_{_\odot}$ and $2\times 10^3\,{\rm AU}$. Next, using a spherical polar coordinate system, $(r,\theta,\phi)$, with the pole along the $z$-axis, the azimuthal angle of each SPH particle, $i$, is changed from $\phi_i$ to $\phi'_i$, where $\phi'_i+0.5\sin(\phi'_i)\cos(\phi'_i)=\phi_i$. Finally, each SPH particle is given an initial velocity ${\bf v}_i=\Omega_{_{\rm O}}\hat{\bf e}_z\!\times\!{\bf r}_i$. When the test is performed with \USs, the results are chaotic, and the results show no clear trends. Since the Boss-Bodenheimer test is intended to illustrate the improved stability of the new procedures for sink {\it evolution}, the results with \USsg are not discussed further.

\begin{figure}
\begin{center}
\includegraphics[width=86mm,angle=0]{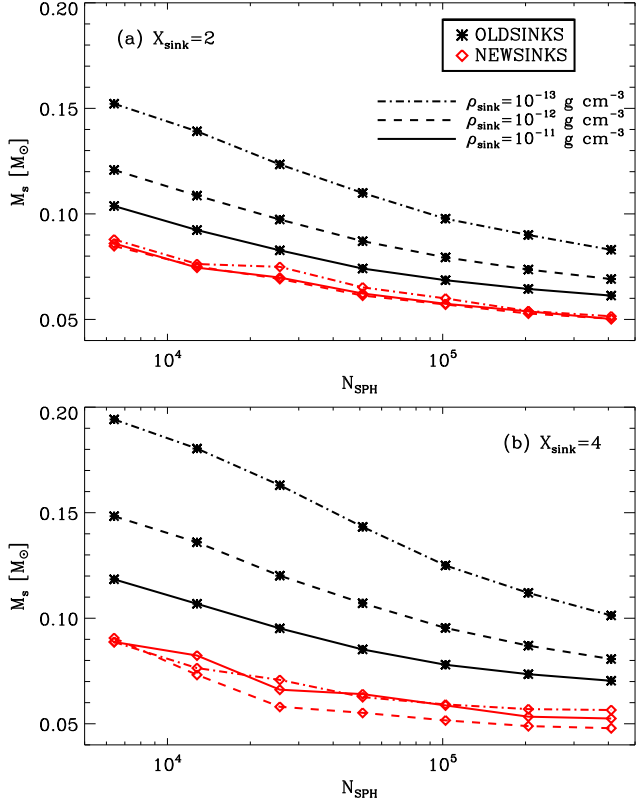}
\caption{The mass of the first sink to form in the Boss-Bodenheimer Test, recorded at time $t\!=\!0.03\,{\rm Myr}$, plotted against the number of SPH particles, ${\cal N}_{_{\rm SPH}}$, for three different sink formation densities $\rho_{_{\rm SINK}}\!=\!10^{-13},\;10^{-12},\;{\rm and}\;10^{-11}\,{\rm g}\,{\rm cm}^{-3}$. Results obtained using \NSsg are shown with red diamonds and red connecting lines, those obtained using \OSsg are shown with black stars and black connecting lines. The upper panel is for $X_{_{\rm SINK}}\!=\!2$ (the default option), and the lower panel is for $X_{_{\rm SINK}}\!=\!4$.}
\label{FIG:BOSSBOD:2}
\end{center}
\end{figure}

The evolution is illustrated on Fig. \ref{FIG:BOSSBOD:1}, which shows the simulation performed with \NSs, $X_{_{\rm SINK}}\!=\!2$, $\rho_{_{\rm SINK}}\!=\!10^{-11}\,{\rm g}\,{\rm cm}^{-3}$ and ${\cal N}_{_{\rm SPH}}\!=\!102400$. As the cloud collapses, it flattens (due to rotation) and at the same time the azimuthal density perturbation is amplified to produce a symmetric binary system with a dense filament between the two components. In Fig. \ref{FIG:BOSSBOD:2} we plot the mass, at time $t=0.030\,{\rm Myr}$, of the first sink to form, against the number of SPH particles, ${\cal N}_{_{\rm SPH}}$, for all the simulations performed. Results obtained with \NSsg are given in red, and those obtained with \OSsg in black. Results obtained with different values of $\rho_{_{\rm SINK}}$ are connected by different line styles. Results obtained with $X_{_{\rm SINK}}=2\;(4)$ are shown in the upper (lower) panel. The second component of the binary system forms almost simultaneously with the first, and its mass at $t=0.030\,{\rm Myr}$ is almost identical to that of the first component, so it is omitted from these plots, to keep them simple.

With \OSs, the results are strongly dependent on numerical resolution, $m_{_{\rm SPH}}={\rm M}_{_\odot}/{\cal N}_{_{\rm SPH}}$, and on the user-defined parameters of sink creation, $\rho_{_{\rm SINK}}$ and $X_{_{\rm SINK}}$. Decreasing $\rho_{_{\rm SINK}}$ causes sinks to form earlier. Increasing  $X_{_{\rm SINK}}$ and/or decreasing $\rho_{_{\rm SINK}}$ and/or decreasing ${\cal N}_{_{\rm SPH}}$ makes sinks larger (increases $R_s$, see Eqn. \ref{EQN:SINKRAD:3}), and this makes the artifical enhancement of the accretion rate greater (see below for the explanation). Both effects (earlier formation, greater artificial enhancement of the accretion rate) increase the mass of the \OSg by $0.030\,{\rm Myr}$.

With \NSs, there is almost no dependence on $\rho_{_{\rm SINK}}$ or $X_{_{\rm SINK}}$, and provided the minimum Jeans mass is well resolved (${\cal N}_{_{\rm SPH}}\ga 10^5$), almost no dependence on ${\cal N}_{_{\rm SPH}}$.

The reasons why increasing $R_s$ amplifies the artificial enhancement of the accretion rate onto an \OSg are threefold. First, to be accreted by an \OSg an SPH particle has to enter its exclusion-zone, which is evidently easier if the exclusion-zone is larger. Second, the flow of SPH particles into the exclusion-zone is amplified artificially by the steep non-physical gradients at $R_s$. Third, once an SPH particle is accreted, it takes with it specific angular momentum of order $(GM_sR_s)^{1/2}$. The larger $R_s$ is, the larger the amount of angular momentum that is removed by accretion, when this angular momentum should be transferred to the matter that is next in line to be accreted, thereby reducing its chances of being accreted in the immediate future.

With \NSs, this does not happen. First, the flow of SPH particles into the interaction-zone is not amplified artificially by steep non-physical gradients at $R_s$. Second, the SPH particles that are accreted by the point-mass have normally transferred much more angular momentum (to other SPH particles) in the process of migrating into the centre of the interaction-zone. Third, whatever angular momentum they do add to the point-mass is then returned to the SPH particles remaining in the interaction-zone, so that the angular momentum of the point-mass remains small.

\begin{figure*} 
\begin{center}
\includegraphics[width=170mm,angle=0]{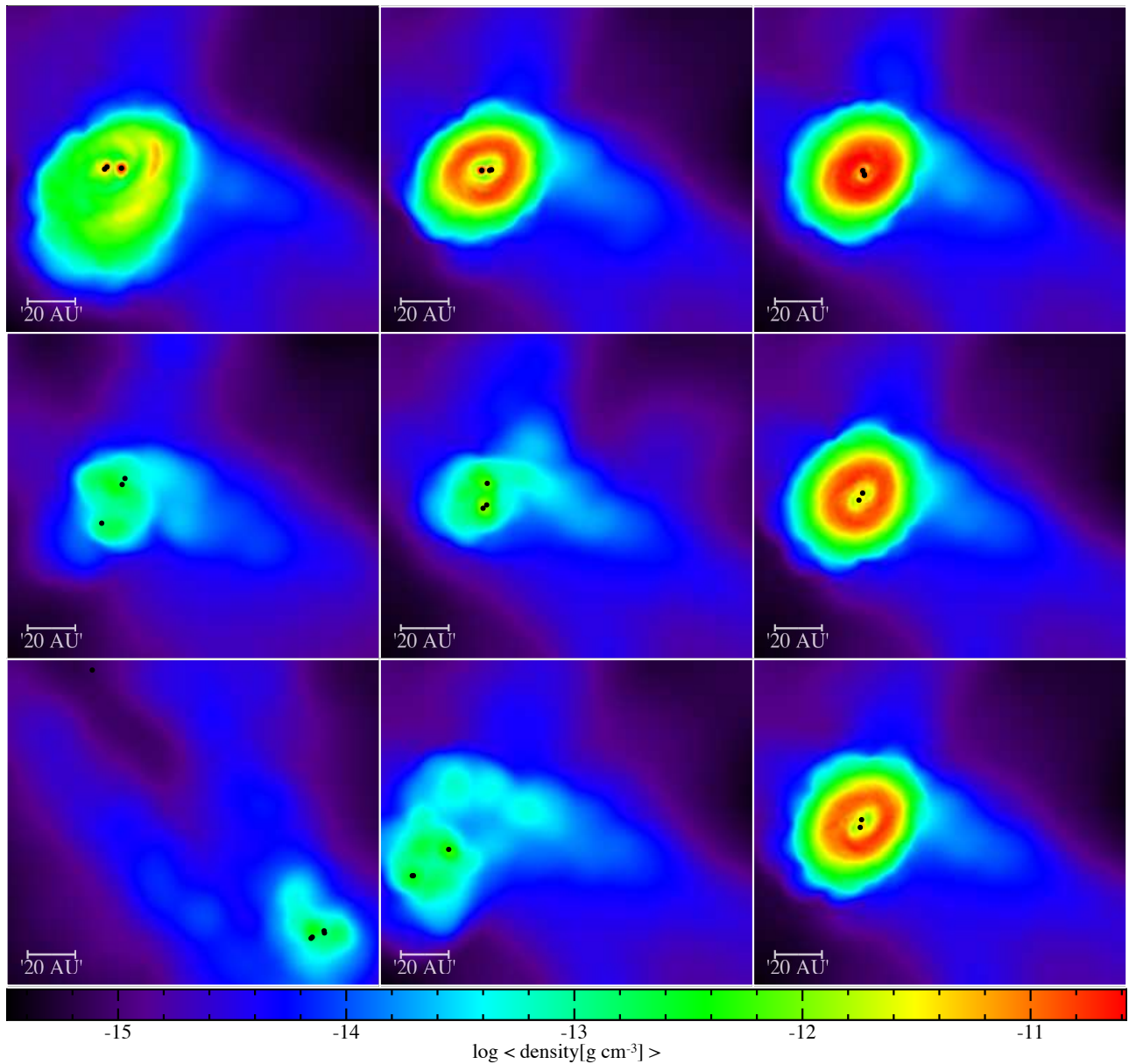}
\caption{False-colour images of $\log_{_{10}}(\bar{\rho}/{\rm g}\,{\rm cm}^{-3})$ at time $t_{_{\rm FIN}}\!=\!150\;{\rm kyr}$ for the simulations of a collapsing turbulent prestellar core; $\bar{\rho}$ is the density-weighted mean density along the line of sight, and black dots mark sinks.  The upper row shows frames from simulations performed with \NSs, middle row \OSs, and bottom row \USs. From left to right the three columns show results obtained with $\rho_{_{\rm SINK}}=10^{-11},\;10^{-10},\;{\rm and}\;10^{-9}\,{\rm g}\,{\rm cm}^{-3}$.}
\label{FIG:TURBCORE}
\end{center}
\end{figure*}

\begin{table*}
\begin{center}
\begin{tabular}{c|c|ccc|c|ccc|c|ccc}\hline
$\underline{\rho_{_{\rm SINK}}}$ & &            &            &           & &            &            &           &
 &            &            &           \\
$({\rm g}\,{\rm cm}^{-3})$       & & $10^{-11}$ & $10^{-10}$ & $10^{-9}$ & & $10^{-11}$ & $10^{-10}$ & $10^{-9}$ &
 & $10^{-11}$ & $10^{-10}$ & $10^{-9}$ \\\hline
ID  & $\,$ & \multicolumn{3}{c}{$\longleftarrow\;\;$\NSs$\;\;\longrightarrow$} 
    & $\;$ & \multicolumn{3}{c}{$\longleftarrow\;\;$\OSs$\;\;\longrightarrow$}
    & $\;$ & \multicolumn{3}{c}{$\longleftarrow\;\;$\USs$\;\;\longrightarrow$} \\
$1$ & & $86.8,0.12$ & $87.5,0.11$ & $89.5,0.10$ & & $86.8,0.16$ & $87.5,0.14$ & $89.5,0.19$ &
 & $86.8,0.13$ & $87.5,0.20$ & $89.5,0.20$ \\
$2$ & & $87.9,0.12$ & $89.2,0.11$ & $91.2,0.10$ & & $88.1,0.14$ & $89.3,0.22$ & $91.2,0.18$ &
 & $87.9,0.11$ & $89.2,0.13$ & $113,0.18$  \\
$3$ & & $110,0.19$  & $112,0.16$  & $113,0.14$  & & $110,0.18$  & $112,0.12$ & \dsh \dsh   &
 & $110,0.08$  & $112,0.11$  & \dsh \dsh   \\
$4$ & & \dsh \dsh   & \dsh \dsh   & \dsh \dsh   & & \dsh \dsh   & \dsh \dsh   & \dsh \dsh   &
 & $113,0.07$  & $125,0.04$  & \dsh \dsh   \\
$5$ & & \dsh \dsh   & \dsh \dsh   & \dsh \dsh   & & \dsh \dsh   & \dsh \dsh   & \dsh \dsh   &
 & $114,0.05$  & \dsh \dsh   & \dsh \dsh   \\
$6$ & & \dsh \dsh   & \dsh \dsh   & \dsh \dsh   & & \dsh \dsh   & \dsh \dsh   & \dsh \dsh   &
 & $114,0.03$  & \dsh \dsh   & \dsh \dsh   \\
$7$ & & \dsh \dsh   & \dsh \dsh   & \dsh \dsh   & & \dsh \dsh   & \dsh \dsh   & \dsh \dsh   &
 & $116,0.07$  & \dsh \dsh   & \dsh \dsh   \\
$8$ & & \dsh \dsh   & \dsh \dsh   & \dsh \dsh   & & \dsh \dsh   & \dsh \dsh   & \dsh \dsh   &
 & $116,0.05$  & \dsh \dsh   & \dsh \dsh   \\
$9$ & & \dsh \dsh   & \dsh \dsh   & \dsh \dsh   & & \dsh \dsh   & \dsh \dsh   & \dsh \dsh   &
 & $131,0.03$  & \dsh \dsh   & \dsh \dsh   \\\hline
$M_{_{\rm TOT}}$ & & $0.43$ & $0.38$ & $0.34$ & & $0.48$ & $0.48$ & $0.37$ & & $0.62$ & $0.48$ & $0.38$ \\
${\cal N}_{_{\rm TOT}}$ & & 3 & 3 & 3 & & 3 & 3 & 2 & & 9 & 4 & 2 \\\hline
\end{tabular}
\end{center}
\caption{Properties of the sinks formed in the nine simulations of a collapsing turbulent prestellar core that are depicted in Fig. \ref{FIG:TURBCORE}. Each column represents a different simulation (different combination of sink algorithm and/or $\rho_{_{\rm SINK}}$), and lists sequentially the formation time, in kyrs, and the mass, in ${\rm M}_{_\odot}$, at $t_{_{\rm FIN}}\!=\!150\,{\rm kyr}$ (thus, for example, in the simulation using \NSsg and $\rho_{_{\rm SINK}}\!=\!10^{-10}\,{\rm g}\,{\rm cm}^{-3}$ the third sink is created after $112\,{\rm kyr}$ and by $150\,{\rm kyr}$ has $M_s\!=\!0.16\,{\rm M}_{_\odot}$). In addition, we give, at the foot of the table, the total mass of sinks formed, $M_{_{\rm TOT}}$, by $t_{_{\rm FIN}}\!=\!150\,{\rm kyr}$, again in ${\rm M}_{_\odot}$, and the total number of sinks, ${\cal N}_{_{\rm TOT}}$.}
\label{TAB:TURBCORE}
\end{table*}

\subsection{Turbulent core collapse} \label{SEC:TURBCORE}%

In order to establish how well \NSsg perform under more realistic conditions, we simulate the collapse of a turbulent prestellar core, using initial conditions from \citet[][specifically core `A-MM8' with seed 200]{Walch2012}. The core has mass $1.28\,{\rm M}_{_\odot}$, initial radius $2.5\times 10^3\,{\rm AU}$, and the density prodile of a critical Bonnor-Ebert Sphere (i.e. one with dimensionless boundary radius $\xi_{_{\rm B}}=6.45$). The initial turbulent velocity field subscribes to a power spectrum of the form $P(k)\propto k^{-4}$ \citep[see also][]{Walch2010, Walch2012}, with $2\!\le\!k\!\le\!10$ (where $k\!=\!1$ corresponds to the core radius). The gas is initially isothermal at $11\,{\rm K}$, {\refrpt and the equation of state is evolved with the algorithm of \citet{SWBG2007}, which captures the transport of cooling radiation and the dependence of the opacity on density and temperature.} The Mach Number of the turbulence is 1.5, and the ratios of thermal and turbulent energy to gravitational energy are $\alpha\!=\!0.018$ and $\gamma\!=\!0.040$.

The initial conditions are set up by cutting a sphere of 157,000 SPH particles from a relaxed periodic cube, and scaling the mass and radius so that the inner 128,000 SPH particles have total mass $M_{_{\rm CORE}}\!=\!1.28\,{\rm M}_{_\odot}$ and outer radius $R_{_{\rm CORE}}\!=\!2.5\times 10^3\,{\rm AU}$; thus all SPH particles have mass $m_{_{\rm SPH}}\!=\!10^{-5}\,{\rm M}_{_\odot}$. Next the inner 128,000 SPH particles are stretched to fit the density profile of a critical Bonnor-Ebert Sphere (as in the rotating Bonnor-Ebert Sphere test of Section \ref{SEC:BONNOREB}), and allocated a temperature $T\!=\!11\,{\rm K}$. At the same time, the outer 29,000 SPH particles are stretched so that they have uniform number-density ten times smaller than the number-density of the SPH particles just inside $R_{_{\rm CORE}}$, and allocated a temperature $T\!=\!110\,{\rm K}$. Finally, the turbulent velocity field is evaluated on a Cartesian grid, and the velocities of individual SPH particles are obtained by interpolation on this grid.

The evolution of the core is simulated using \NSs, \OSsg and \USs, with three different values of $\rho_{_{\rm SINK}}\!=\!10^{-11}$, $10^{-10}$ and $10^{-9}\textrm{ g cm}^{-3}$; {\refrpt with these high values of $\rho_{_{\rm SINK}}$ the gas is well into the adiabatic heating regime by the time sinks are created}. In all cases we set $X_{_{\rm SINK}}\!=\!2$, and all nine simulations are terminated at $t_{_{\rm FIN}}\!=\!150\,{\rm kyr}$. Fig. \ref{FIG:TURBCORE} shows the density field on the $z\!=\!0$ plane at $t_{_{\rm FIN}}$, and the projected positions of the sinks. Table \ref{TAB:TURBCORE} gives the times at which the sinks form and their masses at $t_{_{\rm FIN}}$, plus the total mass and number of sinks at $t_{_{\rm FIN}}$.

The simulations performed with \NSsg appear to be well converged. {\refrpt (i) For all three values of $\rho_{_{\rm SINK}}$, three sinks are formed. (ii) All three sinks are formed at the same time, apart from a small systematic shift that is attributable to the fact that as $\rho_{_{\rm SINK}}$ increases, it takes a little longer for the protostar to contract to $\rho_{_{\rm SINK}}$. (iii) All three sinks are formed in the same place. (iv) The masses of all three sinks at $t_{_{\rm FIN}}$ are the same, apart from a small systematic shift which is attributable to the fact that as $\rho_{_{\rm SINK}}$ increases the size of the sink decreases, and therefore more of the mass destined for the protostar is still in the form of active SPH particles.} The positions of the sinks and the distribution of the residual gas are also essentially the same, again, modulo the fact that there is more residual gas that has not yet been accreted, when $\rho_{_{\rm SINK}}=10^{-9}\,{\rm g}\,{\rm cm}^{-3}$, than when $\rho_{_{\rm SINK}}=10^{-11}\,{\rm g}\,{\rm cm}^{-3}$.

The simulations performed with \OSsg are not converged. Only the times of formation of the first two sinks are more-or-less independent of $\rho_{_{\rm SINK}}$. For $\rho_{_{\rm SINK}}\!=\!10^{-9}\,{\rm g}\,{\rm cm}^{-3}$ the third sink does not form at all (at least, not by $150\,{\rm kyr}$), and in all cases the masses at $150\,{\rm kyr}$ do not vary with $\rho_{_{\rm SINK}}$ in a systematic way.

With \USsg, the situation is chaotic, in the sense that -- apart from the time of formation of the first sink, which is independent of the sink type -- there is no pattern to the changes in creation time and final mass produced by changing $\rho_{_{\rm SINK}}$. The only systematic trend is that the total number of sinks formed increases with decreasing $\rho_{_{\rm SINK}}$. The positions of the sinks, and the distribution of the residual gas are also very different for different values of $\rho_{_{\rm SINK}}$. In short, the simulations with \USsg are divergent.

\section{Discussion}\label{SEC:DISC}%

All three types of sink use the density threshold and overlap criteria (Eqns. \ref{EQN:RHOCRIT} \& \ref{EQN:VRLPCRIT}) for sink creation. \NSsg and \OSsg additionally use the potential minimum and Hill Sphere criteria (Eqns. \ref{EQN:POTMINCRIT} \& \ref{EQN:HILLCRIT}), whereas \USsg use the acceleration divergence, velocity divergence, and non-thermal energy criteria (Eqns. \ref{EQN:DIVACCCRIT}, \ref{EQN:DIVVELCRIT} \& \ref{EQN:ENERGYCRIT}). As a consequence, \USsg are prone to artificial sink creation.

A \NSg comprises a point-mass and a concentric spherical interaction-zone populated by SPH particles; these SPH particles are accreted by the point mass on an extended timescale (regulated accretion) and the angular momentum that they add to the point-mass is returned to the SPH particles remaining in the interaction zone, again on an extended timescale. In contrast, an \OSg or \USg comprises a point-mass and a concentric spherical exclusion-zone that usually is almost devoid of SPH particles. Most SPH particles entering the exclusion-zone are immediately assimilated by the point-mass (instantaneous accretion), taking their angular momentum with them, and this results in artificially steep gradients in the vicinity of the exclusion-zone boundary, {\it and} artificial removal of angular momentum. 

The potential minimum criterion was introduced by \citet{Federrath2010}, and is effective because the gravitational potential derives from all the matter in the computational domain, and therefore is not strongly influenced by particle noise. In combination with the Hill Sphere criterion, it stops density peaks that are destined to be sheared apart by some external gravitational field from forming sinks.

Regulated accretion is designed to cure the problems that arise from the lack of SPH particles in the exclusion zone of a standard sink (\OSg or \US) and the non-physical flow of angular momentum into a standard sink.

The tests we have implemented here are not intended to be exhaustive, but to give an indication of the improvement in performance that \NSsg deliver. First, \NSsg ensure relatively well-behaved hydrodynamics at the sink boundary. Second, they do not act as sinks of angular momentum. Third, as a consequence of these first two features, they predict approximately converged results (i.e. creation times, masses, accretion rates, etc.) that are more-or-less independent of the user-defined sink parameters.

We stress that convergence can only be tested using structures that develop from prescribed perturbations, i.e. perturbations that can be accurately reproduced in the initial conditions of simulations having different resolutions and/or different sink parameters. Thus, for example, in the Boss-Bodenheimer Test, when the two original protostars and their accretion discs interact violently at periastron ($\sim\!0.032\,{\rm Myr}$) several new sinks may form. However, these have their origin in fluctuations due to particle noise, and therefore their genesis is a chaotic process that cannot be used as the basis of a convergence test. It is for this reason that we chose to measure the mass of the first protostar to form, and to make this measurement at $\sim\!0.030\,{\rm Myr}$.

\NSsg incur a significant computational overhead, but, given the non-convergence and non-physicality inherent in the use of standard sinks, this overhead is a price that has to be paid. For example, in the turbulent core collapse test of Section \ref{SEC:TURBCORE}, switching from \OSsg to \NSsg increases the number of CPU-hours from 68 to 253 when $\rho_{_{\rm SINK}}\!=\!10^{-11}\,{\rm g}\,{\rm cm}^{-3}$, and from 313 to 766 when $\rho_{_{\rm SINK}}\!=\!10^{-9}\,{\rm g}\,{\rm cm}^{-3}$. The increase in CPU time derives mainly from the fact that the SPH particles inside the interaction-zone are -- of necessity -- evolved with very short timesteps.

A further advantage of \NSsg is that, since they produce converged -- and therefore hopefully accurate -- accretion rates, one can have more confidence in using these to generate accretion luminosities (radiative feedback) and outflows (mechanical feedback) from protostars or black holes. Sub-grid physics can easily be added to a \NS, for example a model of stellar evolution with accretion \citep[cf.][]{Krumholz2007}, a model for episodic accretion \citep[cf.][]{Stametal2011}, or a model with mechanical feedback \citep[cf.][]{Stametal2005, Cunningham2011}.

The \NSg algorithm has some similarities to the scheme used by \citet{Springel2005} to model accretion onto black holes in galactic nuclei. This scheme estimates an accretion rate using the Bondi-Hoyle-Littleton theory \citep{Hoyle1939, Bondi1944, Bondi1952} and the local density, sound speed and velocity dispersion; the SPH particles to be accreted are chosen stochastically, but with a weighting that preferentially selects those nearest the BH; and the dynamical mass of the BH is increased smoothly so that the accretion of small numbers of massive SPH particles does not cause large feedback fluctuations. We have also experimented with using an average of the Bondi accretion rates for the SPH particles in the interaction-zone to determine the time-scale for radial accretion, but we find that Eqn. (\ref{EQN:tRAD}) is better behaved. Moreover, it may sometimes be more realistic to treat accretion onto a BH in a galactic nucleus using the disc-accretion limit (as would happen automatically with a \NS). Additionally, in the \citet{Springel2005} scheme the BH assimilates all the angular momentum of an accreted SPH particle, which may lead to unrealistic effects.

{\refrpt We stress that the issues of robust sink creation that we have addressed here are quite distinct from those of resolution, which we have not addressed. For example, if, as in our code, inter-particle gravity is kernel-softened, the minimum Jeans mass is only adequately resolved if it contains very many more particles than the mean number of neighbours (i.e. a large multiple of $\bar{\cal N}_{_{\rm NEIB}}$), so that gravitational interactions between the majority of particles are not softenned; that is a resolution issue. However, if sinks are not created and evolved in such a way as to avoid spurious creation, unphysical boundary gradients, and unphysical assimilation of angular momentum, then, even if the self-gravitating gas dynamics that leads to sink creation and subsequently feeds sinks is faithfully simulated, with high resolution, the properties of the sinks formed are not converged, and cannot be trusted; these are the issues that \NSsg deal with.}

\section{Conclusions} \label{SEC:CONC}%

We have developed a new algorithm for the creation and evolution of sink particles in SPH, which has four user-defined free parameters, $(\rho_{_{\rm SINK}},\,X_{_{\rm SINK}},\,X_{_{\rm HILL}},\,\alpha_{_{\rm SS}}$), with default values ($10^{-11}\,{\rm g}\,{\rm cm}^{-3},\,2,\,4,\,0.01$). This algorithm has several advantages over the standard algorithms used heretofore in SPH simulations of star formation. (i) SPH particles flow into the volume of a \NSg (its interaction-zone), and are only accreted by the point-mass later, piecemeal, and usually over many timesteps. Consequently the hydrodynamics in the vicinity of the boundary of a \NSg is not so severely corrupted by poorly evaluated gradients. (ii) The angular momentum of SPH particles assimilated by the point-mass is transferred back to the SPH particles in the interaction-zone, so that the point-mass does not act as a sink for angular momentum; this angular momentum is in any case much reduced by the fact that the accreted SPH particles have normally had to lose angular momentum to other SPH particles before they become the closest SPH particle to the point-mass (and therefore next in line to be accreted). These factors mean that the rate of accretion onto a \NSg is both better informed by physics than that onto a standard sink, {\it and} relatively independent of the user-defined parameters of sink creation. Provided the simulation has sufficient resolution, well converged results are obtained up to density $\rho_{_{\rm SINK}}$. We have measured the new algorithm against four tests: Bondi accretion, the collapse of a uniformly rotating Bonnor-Ebert sphere, the Boss-Bodenheimer Test, and the collapse of a turbulent prestellar core. In all cases the \NSsg appear to produce quite reliable results (i.e. accurate where there is a known solution, and otherwise approximately converged), whereas the standard algorithms do not. If $\rho_{_{\rm SINK}}$ is fixed, then a simulation with \NSsg requires a factor between two and four times more computing resource than one with standard sinks,. However, this expense must be set against the fact that results obtained with \NSsg are more credible. Furthermore, since the results obtained with \NSsg are converged at lower values of $\rho_{_{\rm SINK}}$, one can safely reduce the computing time -- by a comparable factor -- by {\refrpt decreasing} $\rho_{_{\rm SINK}}$.

\section*{Acknowledgements}%

DAH is funded by a Leverhulme Trust Research Project Grant (F/00 118/BJ) and an STFC post-doc. SW gratefully acknowledges the support of the DFG Priority Programme No. 1573. APW gratefully acknowledges the support of the STFC rolling grant PP/E000967/1. APW and SW acknowledge the support of the Marie Curie CONSTELLATION Research Training Network. Most of the simulations were performed on the Cardiff University ARCCA Cluster. {\refrpt We thank the referee, Daniel Price, and also Christoph Federrath, for constructive comments that helped to improve an earlier version of this paper.}  We thank Matthew Bate for helpfully clarifying some details of the original sink particle implementation.  We again thank Daniel Price for supplying the SPLASH code \citep{Price2007}, which was used to prepare Figs. \ref{FIG:BOSSBOD:1} and \ref{FIG:TURBCORE}



\end{document}